\documentclass[fleqn,usenatbib]{mnras}
\usepackage{newtxtext,newtxmath}

\usepackage[T1]{fontenc}

\DeclareRobustCommand{\VAN}[3]{#2}
\let\VANthebibliography\thebibliography
\def\thebibliography{\DeclareRobustCommand{\VAN}[3]{##3}\VANthebibliography}

\usepackage{anyfontsize}
\usepackage{graphicx}	
\usepackage{amsmath}	
\usepackage{orcidlink}
\usepackage[flushleft]{threeparttable}
\usepackage{array}
\usepackage[normalem]{ulem}
\newcommand{\aref}[1]{\hyperref[#1]{Appendix~\ref*{#1}}}

\usepackage{xcolor}

\newcommand{\km}{{\rm{km}}}
\newcommand{\au}{{\rm{au}}}
\newcommand{\gcc}{{{\rm g\ cm^{-3}}}}

\newcommand{\Jkg}{{{\rm J\ kg^{-1}}}}

\newcommand{\kb}{{k_{\rm{B}}}}
\newcommand{\mH}{{m_{\rm{H}}}}

\newcommand{\Mtot}{{M_{\rm{tot}}}}
\newcommand{\Mdisk}{{M_{\rm{d}}}}
\newcommand{\Mdiski}{{M_{\rm{d0}}}}
\newcommand{\Mstar}{{M_{\rm{*}}}}
\newcommand{\Lstar}{{L_{\rm{*}}}}
\newcommand{\Msun}{{M_{\sun}}}
\newcommand{\Lsun}{{L_{\sun}}}

\newcommand{\ap}{{a_{\rm{p}}}}
\newcommand{\Mp}{{m_{\rm{p}}}}
\newcommand{\Rp}{{r_{\rm{p}}}}
\newcommand{\rhop}{{\rho_{\rm{p}}}}
\newcommand{\rhobase}{{\rho_{\rm{b}}}}
\newcommand{\matm}{{m_{\rm{atm}}}}
\newcommand{\matmi}{{m_{\rm{atm0}}}}
\newcommand{\Mearth}{{M_{\earth}}}
\newcommand{\matmearth}{{m_{\rm{atm\earth}}}}
\newcommand{\matmvenus}{{m_{\rm{atm,Venus}}}}

\newcommand{\fatm}{{f_{\rm{atm}}}}

\newcommand{\fatmearth}{{f_{\rm{atm\earth}}}}

\newcommand{\fescv}{{f_{\rm{esc,v}}}}
\newcommand{\fesc}{{f_{\rm{esc}}}}
\newcommand{\Dmin}{{D_{\rm{min}}}}
\newcommand{\Dmax}{{D_{\rm{max}}}}
\newcommand{\Dvc}{{D_{\rm{vc}}}}
\newcommand{\Dim}{{D_{\rm{im}}}}
\newcommand{\Dbl}{{D_{\rm{bl}}}}
\newcommand{\rhoimp}{{\rho_{\rm{imp}}}}
\newcommand{\mimp}{{m_{\rm{imp}}}}
\newcommand{\macc}{{m_{\rm{acc}}}}
\newcommand{\macctot}{{m_{\rm{acc,tot}}}}

\newcommand{\Rcol}{{R_{\rm{col}}}}
\newcommand{\Rml}{{R_{\rm{ml}}}}
\newcommand{\Rmlcr}{{R_{\rm{ml,cr}}}}
\newcommand{\Rmlcc}{{R_{\rm{ml,cc}}}}
\newcommand{\Rmlccone}{{R_{\rm{ml,cc,1}}}}
\newcommand{\Rmlcctwo}{{R_{\rm{ml,cc,2}}}}

\newcommand{\tauc}{{\tau_{\rm{c}}}}
\newcommand{\tauci}{{\tau_{\rm{c0}}}}
\newcommand{\tauacc}{{\tau_{\rm{acc}}}}

\newcommand{\vimp}{{v_{\rm{imp}}}}
\newcommand{\vrel}{{v_{\rm{rel}}}}
\newcommand{\vesc}{{v_{\rm{esc}}}}

\newcommand{\sigv}{{\sigma_{\rm{v}}}}

\newcommand{\flr}{{f_{\rm{lr}}}}

\newcommand{\Qdstar}{{Q_{\rm{D}}^*}}
\newcommand{\Xc}{{X_{\rm{c}}}}

\title[Giant Impact Ejecta Driving Atmospheric Loss]{Re-accretion of Giant Impact Ejecta Can Drive Significant Atmospheric Erosion on Terrestrial Planets}

\author[T. Ghosh et al.]{
Tuhin Ghosh \orcidlink{0000-0002-3103-2000}$^{1}$\thanks{E-mail:\href{mailto:tghosh.astro@gmail.com}{tghosh.astro@gmail.com}},
Mark Wyatt \orcidlink{0000-0001-9064-5598}$^{1}$
and Oliver Shorttle \orcidlink{0000-0002-8713-1446}$^{1,2}$
\\
$^{1}$ Institute of Astronomy, University of Cambridge, Madingley Road, Cambridge CB3 0HA, UK\\
$^{2}$Department of Earth Sciences, University of Cambridge, Downing Street, Cambridge CB2 3EQ, UK\\
}

\date{Accepted XXX. Received YYY; in original form ZZZ}

\pubyear{\the\year{}}

\begin{document}
\label{firstpage}
\pagerange{\pageref{firstpage}--\pageref{lastpage}}
\maketitle

\begin{abstract}
Giant impacts, the collisions between planetary embryos, play a crucial role in sculpting the planets and their orbital architectures. Numerical simulations have advanced our understanding of these events, enabling estimations of mass and atmospheric loss during the primary impacts. However, high computational costs have restricted investigations to the immediate aftermath, limiting our understanding of the longer-term consequences. In this study, we investigate the effect of re-accretion of giant impact debris, a process previously overlooked, on the atmospheres of terrestrial planets. Following the collisional and dynamical evolution of the debris ejected during the primary impacts, we quantify the amount of debris that would be re-accreted by the progenitor. We find that $\sim 0.003\ \Mearth$ would be re-accreted over a wide range of Earth-like planet properties, assuming $1\%$ of their mass is ejected as non-vaporised debris. Over a prolonged period, the secondary impacts during re-accretion drive enhanced atmospheric loss. Notably, the impacts from the debris of the canonical Moon-forming impact would have gradually eroded an atmosphere similar to present-day Earth's in $\sim 30$ Myr. More generally, any planet growing via giant impacts within $2\ \au$ is likely to experience significant post-impact atmospheric erosion unless the initial atmosphere was at least $5$ times more massive than Earth's. Our results highlight the crucial role secondary impacts from giant-impact ejecta could have in driving the long-term atmospheric evolution of Earth-like planets, and demonstrate that giant impacts can be significantly more effective at eroding such atmospheres than previously thought, when re-accretion of debris is considered.
\end{abstract}



\section{Introduction}

The discovery of thousands of exoplanets in recent decades has revolutionized our understanding of planetary systems beyond our own \citep[e.g.,][]{2015_Winn_Fabrycky, 2021_Zhu_Dong}. With advancements in observational precision, we can now not only discover these planets but also begin to characterize their atmospheres \citep[e.g.,][]{de_Wit_2018, 2020_Madhusudhan, 2024_JWST_atm_roadmap}. These atmospheres serve as windows into the surface conditions, geochemical processes, and the potential habitability of exoplanets \citep[e.g.,][]{madhusudhan_2016, 2020_Kopparapu, Byrne_2024}. However, the present-day atmospheric compositions of planets may differ from that acquired during their formation, as various mechanisms can contribute to both atmospheric growth and erosion over time \citep[e.g.,][]{Zahnle_2017_Cosmic_shoreline}. Therefore, understanding the processes that govern the evolution of planetary atmospheres is essential in interpreting the observations.

Atmospheric growth can occur through processes such as mantle degassing and the delivery of volatiles by comets and asteroids, both of which are thought to have contributed to the formation of Earth's current atmosphere \citep[e.g.,][]{Elkins-Tanton_2008, Lebrun_2013, 2017_Avice, Wyatt_2020, Sinclair_2020}. Conversely, atmospheric erosion, driven by processes including photo-evaporation and/or core-powered mass loss, plays a critical role in sculpting the observed radius distribution of short-period exoplanets \citep[e.g.,][]{2017_Owen_Wu, 2018_Ginzburg_CorePowered}. Giant impacts during the final stage of planetary accretion are also a driver of atmospheric loss \citep[e.g.,][]{2003_Genda_Abe, 2005_Genda_Abe, Schlichting_2015, Schlichting_Mukhopadhyay_2018, 2020_Kegerreis, 2020_Kegerreis_APJL, 2020_Denman, 2022_Denman, 2022_Izidoro_GI_Radius_Valley, Ghosh_Chatterjee_2024, 2024_Lock_Stewart, 2025_Roche}, but their efficiency has been historically underestimated as past studies have primarily considered their immediate effects, neglecting longer-term consequences. In this study, we revisit the efficiency of impact-driven atmospheric loss following giant impacts.

In the final phase of planet formation, Mars-sized planetary embryos grow via colliding among themselves in giant impacts, driven by a chaotic phase of dynamical instabilities, which shapes the orbital and structural properties of the planets we observe today \citep[e.g.,][]{Gabriel_2023}. In our solar system, such impacts are hypothesized to have led to the formation of Earth's Moon \citep[e.g.,][]{Cameron_Ward_1976, Hartmann_1975, 1986_Benz, 2001_Canup, Canup_2012, Cuk_Stewart_2012, Lock_2018} and stripped the mantle of Mercury \citep[e.g.,][]{Benz_1988, Benz_2007, Chau_2018}. In extrasolar systems, it now well-established that instabilities play a key role in shaping the orbital architectures \citep[e.g.,][]{1996_Rasio_Ford, Chatteerjee_2008, 2017Izidoro, Frelikh_2019, Ghosh_Chatterjee_2023, Chen_2025}, with giant impacts contributing to the `peas-in-a-pod' trend observed in the Kepler systems \citep{Ghosh_Chatterjee_2023, 2023_Lammers}. Giant impacts are also thought to be responsible for creating the high bulk density planets known as super-Mercuries \citep[e.g.,][]{Reinhardt_2022, Dou_2024} as well as the density disparities, observed in some of the adjacent planet pairs \citep[e.g.,][]{Liu_2015, Inamdar_2016, Bonomo_2019}.

Most giant impacts between protoplanets do not result in perfect mergers but rather lead to partial accretion and hit-and-run encounters \citep[e.g.,][]{2004_Agnor_Asphaug, 2006_Asphaug_Nature, 2012_Genda, 2012_Leinhardt}. The outcomes of these collisions have been extensively studied through numerical simulations in various contexts, ranging from the canonical Moon-forming impact \citep[e.g.,][]{1986_Benz,2001_Canup, CANUP_2013, Kegerreis_2022} to the collisions between newly discovered super-Earth and sub-Neptune class planets \citep[e.g.,][]{2009_Marcus, 2010_Marcus, Liu_2015, 2017_Hwang, 2018_Hwang, 2020_Denman, 2022_Denman, Ghosh_Chatterjee_2024, 2025_Roche}. Such systematic investigations of giant impacts across a broad parameter space, including variations in impact angle, impact velocity, and the mass and composition of both the impactor and the target, provide valuable insights into the violent final stage of planetary accretion. These simulations constrain the amount of mass ejected during the impact \citep[e.g.,][]{2012_Leinhardt, 2019_Cambioni, 2020_Emsenhuber} and have also led to the development of scaling laws to estimate the extent of atmospheric erosion in such events \citep[e.g.,][]{Shuvalov2009, 2020_Kegerreis_APJL, 2022_Denman}. These studies commonly suggest that only partial atmospheric erosion occurs in typical giant impacts. For instance, the atmospheric mass loss prescription from \citet{2020_Kegerreis_APJL} estimates that only $10\%$ of Earth's primordial atmosphere could have been lost from the immediate effects of the canonical Moon forming impact.

Due to the prohibitive computational cost associated with these numerical simulations, the primary focus of these studies has typically been on the impact and its immediate aftermath, typically covering only a few days. However, these impacts generally eject a fraction of mass from the colliding bodies, some of which persists within the system long after the primary impact \citep[e.g.,][]{Jackson_Wyatt_2012, Jackson_2014, Genda_2015, Watt_2024}.
Depending on the specifics of the impact, some fraction of this ejected mass would be vaporised \citep[e.g.,][]{Lock_Stewart_2017, Carter_2020, 2020_Davies_JGRE, Watt_2021, Nakajima_2022, 2023_Caracas_Stewart}. As the vaporised ejecta expands, it cools and condenses, producing dust at detectable levels, which are then rapidly depleted to non-detectable levels through collisional evolution and subsequently removed from the system via radiation forces \citep[e.g.,][]{Wyatt_2016, Watt_2021}.
The non-vaporised debris survives much longer and will gradually erode through mutual collisions and other physical processes, including gravitational perturbations from the surviving planet(s) \citep[e.g.,][]{Jackson_Wyatt_2012, Watt_2024}. Over time, these objects will be progressively ground down to micrometre-sized dust through collisional evolution and eventually removed by radiation pressure or Poynting-Robertson drag \citep[e.g.,][]{Wyatt_2008, Wyatt_2011}. This theoretical understanding is supported by observations of extrasolar systems with large quantities of warm dust in the terrestrial region, inferred to have been recently created in collision between the planetary embryos \citep[e.g.,][]{Lisse_2009_HD172555, Weinberger_2011, Genda_2015, Leinhardt_2015, Su_2019, Moor_2021, Schneiderman_2021_HD172555, 2023_Kenworthy, Watt_2024, 2025_Su}. Importantly, during this long process, a fraction of the debris will re-impact the planet and be re-accreted into the remaining planet(s) \citep{Jackson_Wyatt_2012, Wyatt_2017}, potentially causing further atmospheric erosion.

The re-accretion of the giant impact debris will lead to a series of smaller impacts on the planet over a much longer timescale than those considered in conventional giant impact studies. While each individual impact may have an insignificant effect, the cumulative effects of a large number of such impacts potentially drive substantial atmospheric erosion since the smaller impactors are known to be highly efficient in eroding atmosphere \citep[e.g.,][]{Shuvalov2009, Schlichting_2015}. Despite their potential significance, the effects of these secondary impacts on planetary atmospheres remain largely unexplored. In this study, we investigate the atmospheric erosion from a post-giant-impact planet due to the re-accretion of the ejected material. We first describe our assumptions about the giant impact debris and the numerical model for their collisional and dynamical evolution in \autoref{sec:setup}; in \autoref{sec:res}, we present our key results focussing on a giant impact such as that hypothesised to have formed the Moon, assess the sensitivity of our results to parameter variations, and further explore the outcomes of giant impacts across a broader parameter space; we discuss the potential caveats and various implications of our results in \autoref{sec:discussion}; and conclude in \autoref{sec:conclusions}.

\section{Methods}\label{sec:setup}

In this study, we adopt a simplified approach to investigate the long-term consequences of giant impacts. Depending on the kinematics, giant impacts can result in a range of different outcomes, from efficient accretion and hit-and-run collisions to catastrophic disruptions \citep[e.g.,][]{2012_Leinhardt, 2020_Emsenhuber}. In all classes of impact we consider, a fraction of the parent body mass becomes unbound and is ejected. This unbound mass is generally a mixture of vapour and solid material\footnote{By solid material, we refer to all the non-vaporised condensed matter, which may eventually cool to become solid debris.} with varying proportions, depending on the specifics of the impact \citep[e.g.,][]{Cuk_Stewart_2012, Lock_Stewart_2017, Carter_2020, 2020_Davies_JGRE, Watt_2021, Nakajima_2022}. However, in this study, our focus is not on the detailed outcomes of individual giant impact events. Instead, we attempt to parametrize the outcome in a simplified manner that is broadly applicable across a range of giant impacts. Hence, we adopt a simplified model in which a giant impact results in the ejection of a fraction $\fesc$ of the surviving planet’s mass $\Mp$ as unbound solid material and a fraction $\fescv$ as vapor. The unbound ejecta would expand out from the point of collision and shear out to eventually form an axisymmetric debris disk around the host star encompassing the planet and the collision point \citep[e.g.,][]{Jackson_Wyatt_2012, Jackson_2014, Genda_2015, Watt_2024}. The debris are assumed to have some velocity dispersion in addition to the Keplerian velocity of the planet.

\subsection{Evolution of Debris from Giant Impacts \& Re-accretion}\label{subsec:debris_evol}

In our model, the debris from giant impacts evolve through two main processes: a series of collisions among themselves and interactions with the remaining planet(s).

The successive mutual collisions between the debris progressively break them into smaller fragments, resulting in a collisional cascade that eventually produces micron-sized dust to be removed by radiation pressure. In a steady-state collisional cascade, the size distribution of debris follows a power-law, where the number of bodies with diameters between $D$ and $D + dD$ scales with $D^{-\alpha}$. We adopt $\alpha = 3.5$, which is appropriate if the dispersal threshold is size-independent \citep[e.g.,][]{Wyatt_2007}. The timescale, $\tauc$, over which the impact debris disk would lose half its mass due to both cratering and catastrophic collisions among this population of debris, is calculated in \aref{app:tau_cc}. This timescale strongly depends on the total mass of the debris disk, $\Mdisk$, and the size of the largest objects, $\Dmax$, and is given by (\autoref{eq:tauc_final_form})\footnote{Our estimate of $\tauc$ in \autoref{eq:tau_c} is $\sim 7$ times greater than that reported in \citet{Wyatt_2017}. This difference is primarily due to different assumptions regarding relative velocities (\autoref{subsec:discuss:v_rel}) and the inclusion of cratering collisions (\aref{app:tau_cc}).},
\begin{align}\label{eq:tau_c}
    \tauc &= 4.7 \times 10^{-11} \left(\frac{\Mstar}{\Msun}\right)^{-1} \left(\frac{\ap}{\au}\right)^{4} \left(\frac{\Mp}{\Mearth}\right)^{-2/9} \left(\frac{\rhop}{\gcc}\right)^{-1/9} \nonumber\\
    & \hspace{50pt} \times \left(\frac{\Dmax}{\km}\right) \left(\frac{\Qdstar}{\Jkg}\right)^{5/6} \left(\frac{\Mdisk}{\Mearth}\right)^{-1} \hspace{5pt}\text{Gyr},
\end{align}
where, $\Mstar$ is the stellar mass, $\ap$ is the semi-major axis, and $\rhop$ is the bulk density of the planet. Throughout the study, we have assumed that the debris has a density of $\rhoimp= 2.7\ \gcc$, and a dispersal threshold $\Qdstar = 10^5\ \Jkg$. The size of the largest objects in the debris ($\Dmax$) from a giant impact is highly uncertain.
The vaporised component of the ejecta would expand and cool to form condensates with sizes ($\Dvc$) in the millimetre to centimetre range \citep[e.g.,][]{Melosh_Vickery_1991, Johnson_Melosh_2014}, while the non-vaporised material may reach up to several hundred kilometres. For instance, \citet{Genda_2015} found that the largest debris fragments\footnote{Here, debris fragments refer to the unbound ejecta, not to the remnant post-impact protoplanet(s).} could be as massive as half the lunar mass, corresponding to a diameter of $\sim 1380\ \km$. More recently, \citet{Watt_2024} found that the debris fragments are generally larger than $\sim 100\ \km$ in diameter. This suggests that the size distribution of the ejecta should consist of two primary components, the smaller vapour condensates and the larger non-vaporised material, with their relative proportions and the sizes of the largest non-vaporised bodies depending on the specific impact conditions. Given the substantial differences in their sizes, we expect each component to set up a collisional cascade and to evolve essentially independently of the other.
The small vapour condensates would have short collisional lifetimes, and are therefore rapidly depleted from the system, typically within a small number of orbits \citep[see Figure 5 of][]{Wyatt_2016}. Due to their small size, these condensates are ineffective in atmospheric erosion \citep[e.g.,][]{Schlichting_2015} and thus would have a negligible influence on the post-impact planet(s) over their short lifetimes. In contrast, the larger non-vaporised debris would have much longer collisional lifetimes (e.g., $\tau_c\sim6$ Myr for $\Dmax = 100\ \km$, considering an Earth-like progenitor at $1\ \au$ and $\Mdisk=0.01\ \Mearth$), increasing the likelihood of re-accretion and potentially influencing the planet's subsequent evolution. Therefore, we exclude the vaporised component from our analysis and, for the remainder of this study, focus exclusively on the non-vaporised debris, adopting $\Mdisk = \fesc \Mp$.
%

In addition to the collisional evolution, the debris will also dynamically interact with the surviving planet. These interactions will result in further collisions between the debris and the planet, leading to the accretion of material onto the planet over a timescale \citep{Jackson_Wyatt_2012, Wyatt_2017}
\begin{equation}\label{eq:tau_acc}
    \tauacc = 0.01 \left(\frac{\Mstar}{\Msun}\right)^{-1} \left(\frac{\ap}{\au}\right)^{4} \left(\frac{\Mp}{\Mearth}\right)^{-1/3} \left(\frac{\rhop}{\gcc}\right)^{5/6} \hspace{5pt}\text{Gyr}.
\end{equation}
Note that \autoref{eq:tau_acc} assumes that the debris goes into an axisymmetric disk, with its radial and vertical width set by the planet's escape velocity. This geometry does not fully account for the increased re-accretion rate near the collision point at early times \citep[][]{Jackson_Wyatt_2012}. However, this period of increased accretion is short-lived, lasting only a small fraction of $\tauacc$ before the disk becomes axisymmetric, and its effect on the overall calculations is negligible.

The rate of mass loss from the debris disk, driven by these two primary mechanisms, can be estimated as,
\begin{align}\label{eq:mdot_debris}
    \Dot{\Mdisk} &= -\Mdisk/\tauacc -\Mdisk/\tauc \nonumber \\
           &= -\Mdisk/\tauacc -\Mdisk^2/(\tauci \Mdiski).
\end{align}
Here the subscript "0" indicates the values at time $t=0$ following the primary giant impact. \autoref{eq:mdot_debris} can be solved to obtain,
\begin{equation}\label{eq:mdisk_sol}
    \Mdisk = \Mdiski [(\eta + 1) e^{t/ \tauacc} - \eta]^{-1},
\end{equation}
where
\begin{align}\label{eq:eta}
    \eta &= \tauacc / \tauci \nonumber \\
         &= 2.13 \times 10^8 \left(\frac{\Mp}{\Mearth}\right)^{8/9} \left(\frac{\rhop}{\gcc}\right)^{17/18} \left(\frac{\Dmax}{\km}\right)^{-1} \nonumber \\
         & \hspace{42pt} \times \left(\frac{\Qdstar}{\Jkg}\right)^{-5/6} \fesc 
\end{align}
is the ratio of the accretion timescale to the collisional timescale at time $t=0$, quantifying the relative importance of these two depletion mechanisms. Utilizing \autoref{eq:mdot_debris} and \autoref{eq:mdisk_sol}, we further obtain the cumulative mass reaccreted onto the planet as a function of time,
\begin{equation}\label{eq:m_reacc}
    \macc = \Mdiski \eta^{-1} \ln(1 + \eta - \eta e^{-t/ \tauacc}).
\end{equation}

For our adopted values of $\Qdstar$ and $\Dmax$, and assuming a Earth-like bulk planetary density of $\rhop = 5.5\ \gcc$, we can express $\eta$ as,
\begin{equation}\label{eq:eta_approx}
    \eta = 726\ \left(\frac{\Mp}{\Mearth}\right)^{8/9} \fesc .
\end{equation}
Furthermore, for $1 \lesssim \eta \lesssim 10$, relevant for Earth-like terrestrial planets with conservative estimates of $\fesc \sim 0.01$, the total mass eventually accreted (\autoref{eq:m_reacc}, $t \to \infty$) can be approximated by fitting a power-law to the $\eta$ dependence of \autoref{eq:m_reacc} (see \aref{app:fit}) as
\begin{equation}\label{eq:m_reacc_tot_approx}
    \macctot \sim 0.003\ \left(\frac{\Mp}{\Mearth}\right)^{0.59} \left (\frac{\fesc}{0.01}\right)^{0.54}\ \Mearth.
\end{equation}
From \autoref{eq:m_reacc_tot_approx}, it is evident that even if the debris from a giant impact is only $1\%$ of the planet's mass, an Earth-mass planet can re-accrete $\sim 0.003\ \Mearth$ of debris, which is $\sim 3500$ times more massive than Earth's present-day atmosphere. The accretion of such substantial amounts of debris can have significant implications for the planet's atmosphere, depending on the size and impact velocity of the debris, as well as the properties of the target atmosphere.

\subsection{Atmospheric Mass Loss from Re-accretion}\label{subsec:atm_erosion}

Following \citet{Wyatt_2020} and \citet{Sinclair_2020}, we assume that planets have isothermal atmospheres with temperatures given by $T = 278 (\Lstar / \Lsun)^{1/4} (\ap/\au)^{-1/2}$ K, where $\Lstar$ is the stellar luminosity. The atmospheric scale height, $H$ is set by its temperature ($T$), the mean molecular weight ($\mu$), and the surface gravity of the planet ($g$) as, $H = \kb T / (\mu \mH g)$, where $\kb$ is the Boltzmann constant and $\mH$ is the mass of Hydrogen. Using our assumption about the atmospheric temperature and expressing $g$ in terms of the planet mass, $\Mp$ and mean density, $\rhop$ we get,
\begin{equation}\label{eq:atm_scale_height}
    H = 7.3 \times 10^5 \left(\frac{\Lstar}{\Lsun}\right)^{1/4} \left(\frac{\ap}{\au}\right)^{-1/2} \left(\frac{\Mp}{\Mearth}\right)^{-1/3} \left(\frac{\rhop}{\gcc}\right)^{-2/3} \mu^{-1}\hspace{5pt} \rm{m},
\end{equation}
%

To quantify the amount of atmosphere lost when the debris from a giant impact is re-accreted onto the planet we follow the prescription from \citet{Shuvalov2009}, where the outcome of an impact from an object of diameter $D$ and density $\rhoimp$, at a velocity $\vimp$ is determined by the dimensionless erosional efficiency factor
\begin{align}\label{eq:erosional_efficiency_factor}
    \xi &= \left(\frac{D}{H}\right)^{3} \left[\left(\frac{\vimp}{\vesc}\right)^{2} - 1\right] \left[\frac{\rhop}{\rhobase (1 + \rhop/\rhoimp)} \right] \nonumber \\
    & = 0.5 \times 10^{-18} \left(\frac{\Lstar}{\Lsun}\right)^{-1/2} \left(\frac{\ap}{\au}\right) \left(\frac{\Mp}{\Mearth}\right)^{4/3} \left(\frac{\rhop}{\gcc}\right)^{5/3} \left(\frac{\matm}{\Mearth}\right)^{-1} \nonumber \\
    & \hspace{10pt} \times \mu^{2} \left(\frac{D}{\text{m}}\right)^{3} \left[\left(\frac{\vimp}{\vesc}\right)^{2} - 1\right] \left(1 + \rhop/\rhoimp \right)^{-1}
\end{align}
where, $\vesc$ is the escape velocity from the planet's surface, $\rhobase$ is the density at the base of the atmosphere, and $\matm$ ($\simeq 4 \pi \Rp^2 H \rhobase$; $\Rp$ being the radius of the planet) is the mass of the atmosphere. The atmospheric mass lost due to a single impactor of mass $\mimp$ is given by
\begin{equation}\label{eq:dm_atm_single}
    \frac{\delta \matm}{\mimp} = \left[\left(\frac{\vimp}{\vesc}\right)^{2} - 1\right] \chi_{\rm{a}},
\end{equation}
where,
\begin{align}\label{eq:chi_a}
    \log \chi_{\rm{a}} & = -6.375 + 5.239 \log \xi - 2.121 (\log \xi)^2 \nonumber\\
                       &\ \ \ \  + 0.397(\log \xi)^3 - 0.037(\log \xi)^4 \nonumber\\
                       &\ \ \ \  + 0.0013(\log \xi)^5 \hspace{27pt}\text{for $\log \xi < 6$,}\\
                       & = 0.4746 - 0.6438\log \xi \hspace{10pt}\text{for $\log \xi \ge 6$}.
\end{align}
Note that the experiments presented in \citet{Shuvalov2009} extend only up to $\xi \sim 10^6$. The expression above for $\log \chi_{\rm{a}}$ for $\xi > 10^6$ was extrapolated by \citet{Wyatt_2020}, in agreement with the results presented in \cite{Schlichting_2015}. Nevertheless, as our primary focus is on smaller impactors from post-giant-impact debris, this prescription remains well-suited.

After the primary giant impact, we expect numerous small impacts on the planet over a relatively long timescale ($\sim \tauacc$). These impactors would have different sizes in line with the power-law size distribution of the disk (\autoref{subsec:debris_evol}). Integrating over the size distribution of the impactors, we obtain the total atmospheric mass loss per unit impactor mass from \autoref{eq:dm_atm_single},
\begin{equation}\label{eq:dm_atm}
    \frac{\Delta \matm}{\macc}= A \left[\left(\frac{\vimp}{\vesc}\right)^{2} - 1\right] \int_{\xi_{\rm{min}}}^{\xi_{\rm{max}}} \xi^{(1-\alpha)/3}\chi_{\rm{a}}d\xi,
\end{equation}
where
\begin{align}\label{eq:dm_atm_A}
    A & = \frac{1}{3} \left(\frac{4 - \alpha}{\Dmax^{4-\alpha} - \Dmin^{4-\alpha}}\right) \left(\frac{\xi}{D^3}\right)^{(\alpha - 4)/3} \hspace{5pt}\text{for $\alpha \neq 4$},\\
      & =\frac{1}{3 \ln(\Dmax/\Dmin)} \hspace{91pt}\text{for $\alpha = 4$},
\end{align}
and $\xi_{\rm{max}}$ ($\xi_{\rm{min}}$) are the erosional efficiency parameters corresponding to the largest (smallest) objects.

In this study, we adopt a single representative value for the impact velocity of the debris for simplicity, acknowledging that the impact velocity may vary among the debris. Numerical studies of the canonical Moon-forming impact indicate that the ejected debris has a velocity dispersion, $\sigv \simeq 0.46 \vesc$ \citep{Jackson_2014, Wyatt_2016}.  Considering this velocity dispersion and accounting for the gravitational focusing, we estimate the impact velocity of the debris as
\begin{equation}\label{eq:v_imp}
    \vimp \simeq \sqrt{\vesc^{2} + \sigv^{2}} \simeq 1.1 \vesc .
\end{equation}
%

As the planet accretes debris over time, the resulting atmospheric erosion leads to evolving atmospheric properties. To capture this dynamic evolution, we model the process over equal logarithmic time bins. In each time bin (spanning from $t_{\rm{i}}$ to $t_{\rm{i+1}}$), we compute the amount of debris mass accreted ($\Delta \macc$) utilizing \autoref{eq:m_reacc} as $\Delta \macc = \macc(t_{\rm{i+1}}) - \macc(t_{\rm{i}})$. We assume the atmospheric properties remain relatively unchanged within each time bin and estimate the amount of atmospheric erosion due to re-accretion utilizing \autoref{eq:dm_atm}. Accounting for the computed atmospheric erosion, the updated atmospheric properties serve as initial conditions in the next time interval, enabling the iterative computation of cumulative atmospheric erosion over time.
Note that we are assuming a fixed size distribution of the debris throughout the evolution. In reality, the size distribution may also evolve over time. Accurately modelling the evolving size distribution of the debris is a complex task outside the scope of this study.

To estimate the amount of atmospheric erosion due to this re-accretion for Earth-like planets, we can also make the following approximate calculation. With our adopted values of $\vimp = 1.1 \vesc$, $\rhop = 5.5 \gcc$, $\rhoimp = 2.7 \gcc$, $\Dmax = 100\ \km$, and assuming $\mu=29$, the total atmospheric mass loss per unit impactor mass can be roughly approximated from the proper numerical calculations, assuming power-law dependencies:
\begin{equation}\label{eq:dm_atm_macc_approx}
    \frac{\Delta \matm}{\macc} \simeq 0.00136\ \left(\frac{\ap}{\au}\right)^{-0.15} \left(\frac{\Mp}{\Mearth}\right)^{-0.19} \left(\frac{\matmi}{\matmearth}\right)^{0.125}.
\end{equation}
\autoref{eq:dm_atm_macc_approx} shows that impactors with a total mass of only $\sim 0.001\ \Mearth$ would be able to erode the Earth's current atmosphere ($\matmearth = 0.85 \times 10^{-6} \Mearth$), consistent with the findings of \citet{Schlichting_2015}. Subsequently, estimating $\macctot$ from \autoref{eq:m_reacc_tot_approx}, we get an estimate for the fractional atmospheric loss,
\begin{equation}\label{eq:dm_atm_matmi_approx}
    \frac{\Delta \matm}{\matmi} \simeq 4.8\ \left(\frac{\Mp}{\Mearth}\right)^{0.4} \left(\frac{\ap}{\au}\right)^{-0.15} \left(\frac{\matmi}{\matmearth}\right)^{-0.875} \left (\frac{\fesc}{0.01}\right)^{0.54} .
\end{equation}
\autoref{eq:dm_atm_matmi_approx} estimates that an Earth-like planet can potentially lose $\sim 5\ \matmearth$ of its atmosphere due to the re-accretion of the debris ejected by a giant impact, even when the total mass of the debris is only $1\%$ of the planet's mass. Furthermore, the total atmospheric loss due to re-accretion has relatively weak sensitivity to $\Mp$, $\ap$ and even $\matmi$. While the fractional atmospheric mass loss depends strongly on $\matmi$, it also exhibits relatively weak sensitivity to other parameters. Therefore, irrespective of the specific details of the giant impact, the re-accretion of the debris can induce substantial atmospheric loss for Earth-like terrestrial planet atmospheres. Note that the \autoref{eq:dm_atm_macc_approx} and \autoref{eq:dm_atm_matmi_approx} provide only approximate estimates, valid within the vicinity of the nominal values of the parameters appearing in their denominators. For a more precise description, it is essential to numerically evolve the atmospheric properties forward in time, accounting for the progressive atmosphere loss due to debris impacts as described earlier.

In the following section, we will investigate the re-accretion of giant impact debris and its role in atmospheric erosion in more detail by numerically integrating \autoref{eq:dm_atm} forward in time, utilizing \autoref{eq:m_reacc} to estimate $\Delta \macc$ in each timestep.

\section{Results}\label{sec:res}

We first consider the possible implications of debris re-accretion on Earth following a giant impact such as that hypothesised to have formed the Moon. This extensively studied event, observationally constrained by the Earth-Moon configuration \citep[e.g.,][]{Cameron_Ward_1976, 2001_Canup, Timpe_2023, Zhou_2024}, is a natural reference point for our study. We note, however, that the exact nature of the impact scenario that led to the formation of the Moon remains uncertain and debated. It is not the purpose of our study to address these, but to adopt one representative scenario as a reference. While we use the giant impact hypothesis of the Moon formation as a reference point, the focus of our work is not restricted to satellite-forming impacts. Instead, we aim to explore the broader role of typically overlooked debris in causing atmospheric erosion through secondary impacts, regardless of whether the primary impact forms a satellite or not. To this end, we subsequently expand our analysis to explore a broader range of parameter space.

\subsection{The Moon-Forming Impact}\label{subsec:moon_forming_impact}

\begin{figure}
\includegraphics[width=\columnwidth]{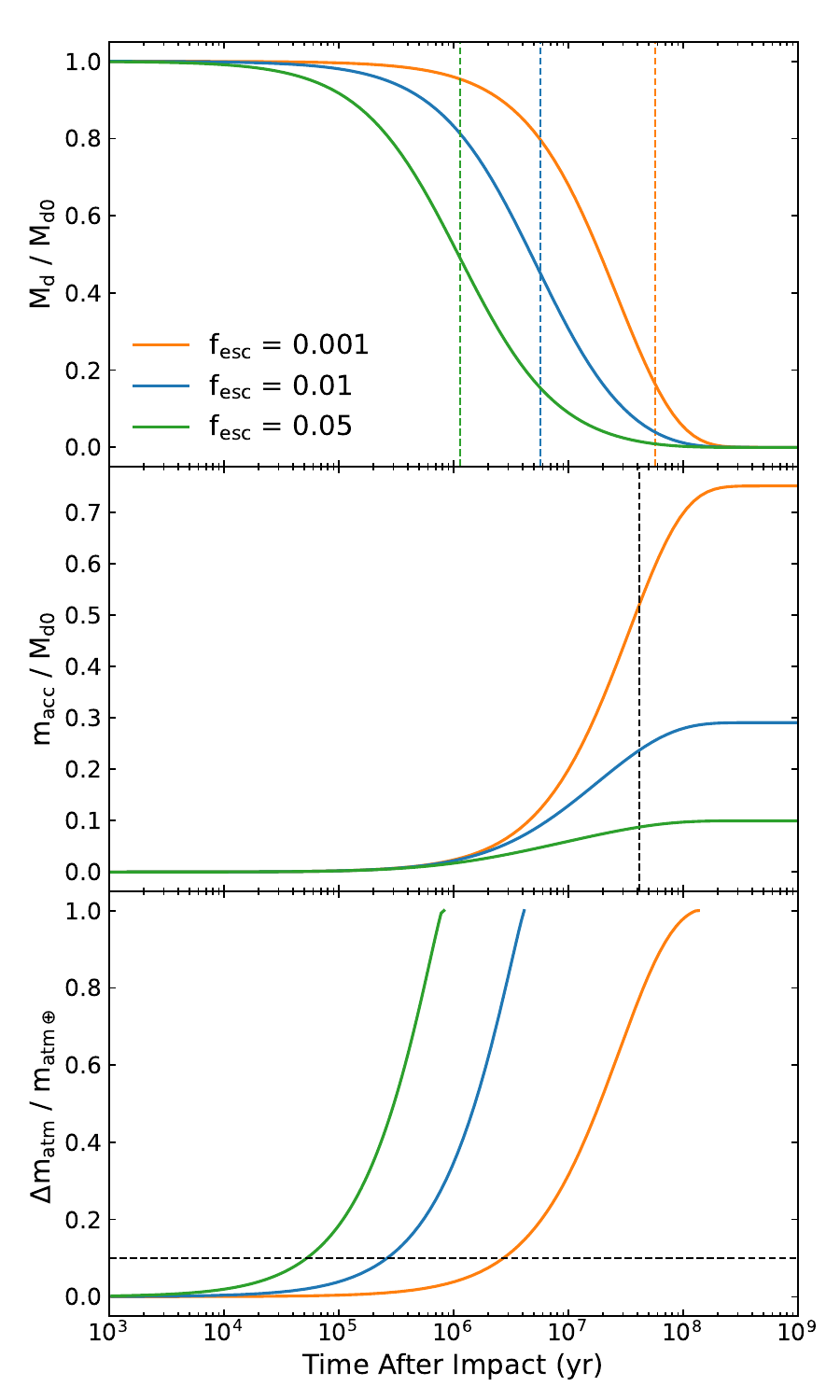}
\caption{
\textit{Top}: Evolution of the ejected debris mass over time following a hypothetical Moon-forming impact with $\Dmax=100\ \km$, according to \autoref{eq:mdisk_sol}. The colors correspond to different values of $\fesc$ (see legend), which determines the initial disk mass, $\Mdiski = \fesc \Mp$. Vertical dashed lines indicate the collisional depletion timescale ($\tauc$, \autoref{eq:tau_c}) for each $\fesc$.
\textit{Middle}: Mass of debris reaccreted onto the planet as a function of time (\autoref{eq:m_reacc}). The black dashed vertical line indicates the re-accretion timescale ($\tauacc$, \autoref{eq:tau_acc}), which does not depend on $\fesc$.
\textit{Bottom}: Atmospheric mass loss over time due to the re-accretion of debris (\autoref{eq:dm_atm}). For reference, the horizontal dashed line represents the atmospheric loss expected from a canonical Moon-forming impact \citep{2020_Kegerreis}.
}
\label{fig:evol_fesc}
\end{figure}

The Moon-forming collision is thought to have been the last major giant impact in the inner solar system. The most extensively studied impact event capable of explaining the angular momentum of the present-day Earth-Moon system is known as the canonical Moon-forming impact. It involved a Mars-sized protoplanet colliding with the proto-Earth at nearly the escape velocity, ejecting a few per cent of Earth's mass as debris \citep[e.g.,][]{Cameron_Ward_1976, 2001_Canup, Canup_2004, Canup_2008}. However, this model has limitations in reproducing all observables \citep[e.g.,][]{Melosh_2014}, which have motivated several alternative hypotheses, including a smaller, high-energy impactor hitting a rapidly spinning proto-Earth \citep{Cuk_Stewart_2012, Lock_Stewart_2017, Lock_2018}, a collision between two protoplanets of comparable mass \citep{Canup_2012}, and a hit and run collision with higher impact velocity \citep{Reufer_2012}. Although no single scenario has been conclusively established as the definitive origin of the Moon, the giant impact hypothesis remains the most widely accepted explanation. Accordingly, in this work, we use the term "Moon-forming impact" to refer broadly to the giant impact event hypothesized to explain the Moon's origin, without implying any specific impact scenario unless stated explicitly. Given the uncertainties in the exact impact scenario and its resultant debris, despite the observational constraints from the Earth-Moon system, our parametric approach is well-suited for the purpose of this study. Our simple parametrization approach is agnostic to the exact kinematics of the giant impact and depends on the velocity distribution of the ejected debris (see \autoref{subsec:discuss:v_rel}), largest fragment sizes ($\Dmax$) and the fraction of the unbound solid mass ($\fesc$), which we treat as variable parameters.

\subsubsection{Effect of $\fesc$}\label{ssubsec:fesc_variation}

Relatively low-energy collisions, such as the canonical scenario \citep[e.g.,][]{2001_Canup}, are expected produce $\fesc \sim 0.01$, whereas high energy \citep[e.g.,][]{Lock_2018} impacts would induce substantial vaporization resulting in significantly reduced $\fesc$. In this section, we systematically vary $\fesc$ and estimate the corresponding atmospheric erosion for each scenario. For this analysis, we assume that immediately following the impact, Earth had orbital and physical properties similar to those of present-day Earth (namely, $\ap = 1\ \au$, $\Mp=\Mearth$, $\rhop = 5.5\ \gcc$, $\matmi = \matmearth = 0.85 \times 10^{-6} \Mearth$, $\mu = 29$) and $\Dmax = 100\ \km$. We have chosen to use Earth's present-day atmosphere as the nominal value due to the lack of constraints of the atmospheric conditions prior to and immediately following the giant impact. We will revisit these assumptions in the following sections.

\autoref{fig:evol_fesc} (top panel) shows the depletion of the debris disk generated from the primary giant impact, the fraction of its mass re-accreted to the planet (middle panel), and the resulting atmospheric erosion (bottom panel) over time for different values of $\fesc$ considered here. In such events involving Earth-like planets, re-accretion typically occurs over a longer timescale than the collisional depletion timescale ($\tauc < \tauacc$, i.e., $\eta < 1$; except for very small $\fesc \lesssim 0.0013$). Consequently, the evolution of the debris disk mass is predominantly governed by collisional depletion. In cases where $\tauc > \tauacc$, re-accretion becomes the dominant debris loss mechanism, leading to the accretion of a greater fraction of mass than that lost by collisional grinding. Since the collisional depletion timescale, $\tauc \propto 1 / \Mdisk$ and $\Mdisk = \fesc \Mp$, as $\fesc$ decreases, the lifetime of the debris disk increases (see \autoref{fig:evol_fesc}, top panel). On the contrary, as $\tauacc$ is independent of $\fesc$ (\autoref{eq:tau_acc}), as $\fesc$ decreases, the planet can re-accrete a larger fraction of the less massive debris disk over its longer lifetime (see \autoref{fig:evol_fesc}, middle panel). However, the total mass re-accreted generally increases with $\fesc$ due to the larger initial disk mass. For reasonable $\fesc$ values, a fraction of the debris inevitably collides with the planet and gets re-accreted within a few $\tauacc$ before the complete collisional depletion of the disk (see \autoref{fig:evol_fesc}, middle panel).

During this period, even with a very conservative estimate of the debris disk mass at $0.1\%$ of Earth's mass ($\fesc = 0.001$), the post-giant-impact Earth accretes $0.00075\ \Mearth$ ($75\%$ of $\Mdiski$) of debris, which is $\sim 885$ times larger than the Earth's atmospheric mass. As a result, the entirety of Earth's atmosphere is lost due to impacts from the ejecta (bottom panel, \autoref{fig:evol_fesc}) within $\sim 135$ Myr. This is particularly noteworthy given that the fractional atmospheric mass loss from the canonical Moon-forming impact is only $\sim10\%$ \citep{Schlichting_2015, 2020_Kegerreis}, assuming $\matmi = \matmearth$ in both cases. As $\fesc$ increases, the planet accretes debris faster, leading to more rapid atmospheric erosion. For $\fesc = 0.01$, Earth accretes $\macc = 0.0029\ \Mearth$ ($29\%$ of $\Mdiski$) , losing its atmosphere in $\sim 4$ Myr. Similarly, for $\fesc = 0.05$, the accreted mass increases to $\macc = 0.005\ \Mearth$ ($10\%$ of $\Mdiski$) , leading to complete atmospheric erosion within $\sim 1$ Myr.

Our findings highlight the extensive atmospheric erosion due to numerous smaller impacts on the collisional remnant as it re-accretes the debris following a giant impact event, leading to complete atmospheric erosions for Earth-like thin atmospheres, even when $\fesc$ is as low as $0.001$. However, for even lower values of $\fesc$ (e.g., in impacts predominantly resulting in vaporization), the total mass of re-accreting solid debris is insufficient to erode the entire atmospheric mass. Although vapour condensates may subsequently re-accrete in such cases, their small sizes and short collisional lifetimes make them largely ineffective in contributing to further atmospheric erosion (\autoref{subsec:debris_evol}).

\subsubsection{Effect of $\Dmax$}\label{ssubsec:dmax_variation}

\begin{figure}
\includegraphics[width=\columnwidth]{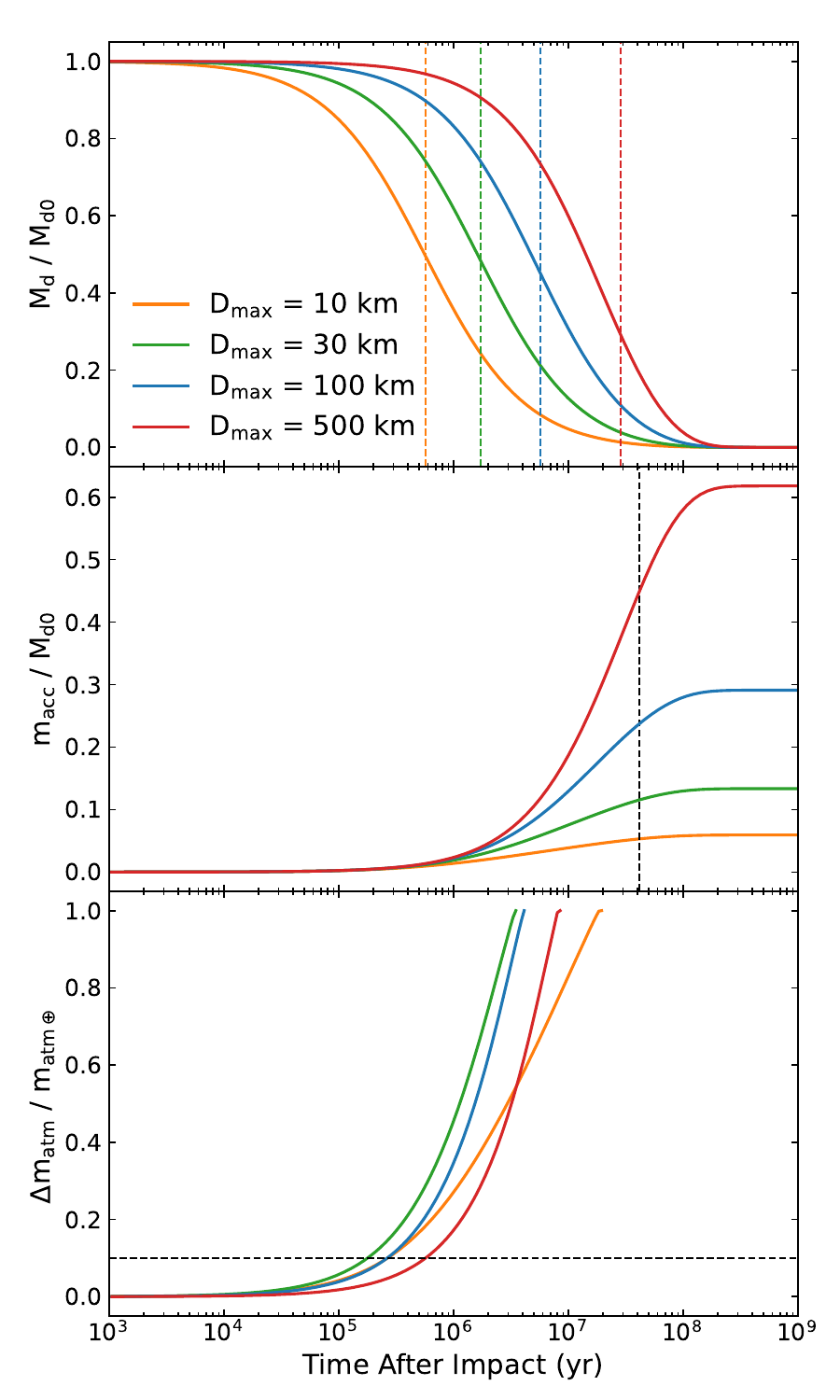}
\caption{
\textit{Top}: Evolution of the debris mass over time (\autoref{eq:mdisk_sol}). The colors correspond to varying values of $\Dmax$ (see legend), assuming $\fesc=0.01$. Vertical dashed lines indicate the collisional depletion timescale ($\tauc$) for each $\Dmax$.
\textit{Middle}: Mass of debris reaccreted onto the planet as a function of time (\autoref{eq:m_reacc}). The black dashed vertical line indicates the re-accretion timescale ($\tauacc$).
\textit{Bottom}: Atmospheric mass loss over time due to the re-accretion of debris (\autoref{eq:dm_atm}). The horizontal dashed line indicates the atmospheric loss expected from a canonical Moon-forming impact \citep{2020_Kegerreis}.
}
\label{fig:evol_dmax}
\end{figure}

In this section, we examine the robustness of our findings against the size of the largest debris, $\Dmax$. In addition to our choice of $\Dmax = 100\ \km$ adopted in the previous section, we explore different values for $\Dmax$, ranging from $5\ \km$ to $500\ \km$. For this analysis, we adopt $\fesc=0.01$\footnote{Note that this choice is not uniquely determined. We selected it as a plausible fiducial value based on typical mass loss fractions inferred in prior studies \citep[e.g.,][]{Jackson_Wyatt_2012, Carter_2020}. But we emphasize that it is not assumed to be universal.}.

The size of the largest objects ($\Dmax$) within the debris disk significantly influences the collisional fragmentation process. Larger objects take more time to fragment into micron-sized dust particles and therefore to have their mass moved into a size fraction affected by radiation forces. Therefore, a disk with larger $\Dmax$ can hold higher mass over a longer lifetime than a disk with a smaller $\Dmax$ (top panel, \autoref{fig:evol_dmax}). This, in turn, leads to both a higher mass accretion rate and a greater total accreted debris mass for larger $\Dmax$ (middle panel, \autoref{fig:evol_dmax}). E.g., for $\Dmax=500\ \km$, the Earth eventually accretes a total debris mass of $\macc = 0.0062\ \Mearth$, whereas for a smaller $\Dmax$ of $30\ \km$ ($10\ \km$) the total accreted debris mass is only $0.0013\ \Mearth$ ($0.0006\ \Mearth$). However, due to the strong dependence of atmospheric mass loss on the size of the impactors, the accretion rates of the debris do not result in proportional atmospheric erosion rates for different $\Dmax$ (bottom panel, \autoref{fig:evol_dmax}). For terrestrial Earth-like atmospheres, the smaller km-sized impactors are more efficient in eroding the atmosphere compared to the larger ones, while sub-km-sized impactors are ineffective \citep[][]{Shuvalov2009, Schlichting_2015}. As the lower values of $\Dmax$ shift the mass distribution of impactors toward this efficient size range, the atmospheric mass loss per unit mass of accreted impactors increases, driving substantial atmospheric losses despite accreting less debris. For $\Dmax=500\ \km$, Earth loses all of its atmosphere within $8.5$ Myr after accreting $0.0016\ \Mearth$ of debris. A lower $\Dmax$ of $30\ \km$ results in faster atmospheric erosion with the complete loss occurring at $3.5$ Myr after accreting only $0.00045\ \Mearth$ of debris. For $\Dmax=10\ \km$, $\tauc \ll \tauacc$, resulting in a much slower mass accretion rate causing complete atmospheric erosion at $19.5$ Myr after accreting $0.00046\ \Mearth$ of debris. At such small $\Dmax$, a significant portion of the debris mass resides in sub-kilometre-sized fragments due to the steep size distribution (\autoref{subsec:debris_evol}). These smaller fragments cannot effectively erode the atmosphere \citep[][]{Shuvalov2009, Schlichting_2015}, which reduces the atmospheric erosion per unit total accreted mass. The combination of decreased erosion efficiency and faster collisional depletion ($\tau_c \propto \Dmax$) causes atmospheric erosion to decline rapidly as $\Dmax$ drops below $10\ \km$. For example, at $\Dmax = 5\ \km$, only $\sim 20\%$ atmosphere is depleted after accreting $0.00034\ \Mearth$ of debris. For even smaller values of $\Dmax$, as well as for the (excluded) vaporised debris component comprising millimetre- to centimetre-sized condensates, the impacts from the small debris would be ineffective at driving significant atmospheric erosion through secondary impacts \citep[e.g.,][]{Schlichting_2015}.

Overall, we find that re-accretion of giant impact debris consistently leads to total atmospheric erosion within $20$ Myr from an Earth-like planet, despite varying $\Dmax$ by a factor of a few around $100\ \km$. These results indicate that the re-accretion of debris likely plays a significant role in post-giant-impact atmospheric erosion, regardless of the specific details of the debris size distribution.

\subsubsection{Effect of Initial Atmospheric Properties}\label{ssubsec:atmos_variation}

In the previous sections, we assumed that Earth, after the giant impact, had an atmosphere similar to that of the present-day Earth. However, the initial atmospheric and surface conditions can have a significant effect on the subsequent atmospheric erosion \citep[e.g.][]{2024_Lock_Stewart}. In this section, we explore the long-term atmospheric erosion following the primary impact, focusing on the effects of initial atmospheric mass and mean molecular weight on this process.

\begin{figure}
\includegraphics[width=\columnwidth]{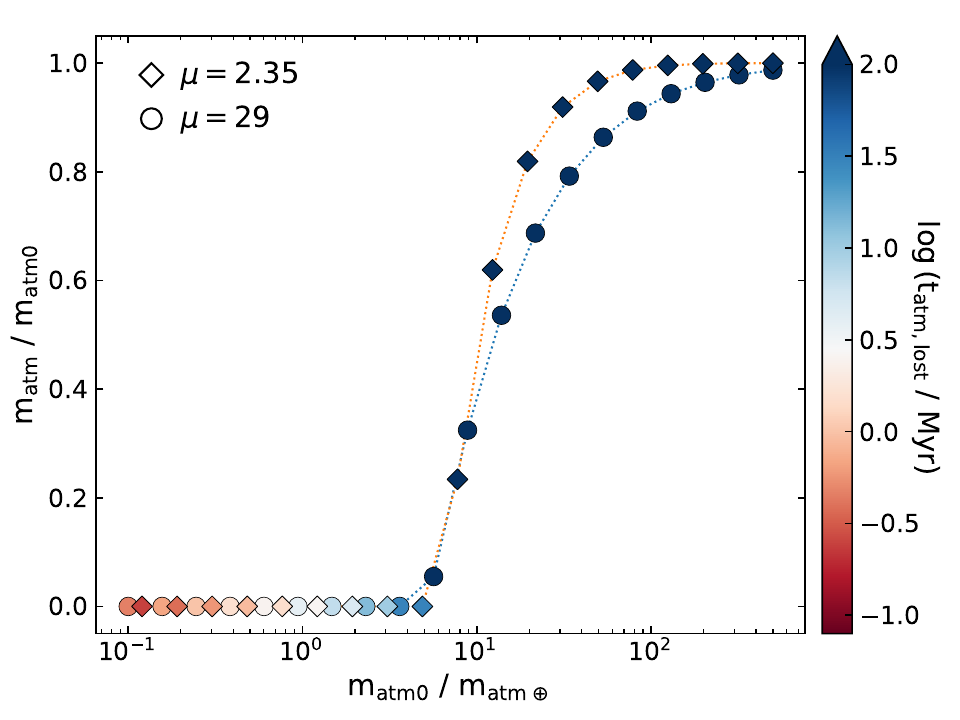}
\caption{Fraction of atmosphere retained $1$ Gyr after a hypothetical Moon-forming impact with $\fesc = 0.01$ and $\Dmax=100\ \km$, as a function of the initial atmospheric mass ($\matmi$). Filled circles represent an Earth-like, volatile-rich atmosphere with $\mu = 29$, and the filled diamonds represent an atmosphere with a primordial (solar) composition with $\mu = 2.35$. For completely depleted atmospheres ($\matm/\matmi < 10^{-4}$), the time of depletion ($t_{\rm{atm,lost}}$) is indicated by the fill colors. Dotted lines connecting markers of the same $\mu$ values are included as visual guides.}
\label{fig:f_atm_m_atm_mu}
\end{figure}

The mean molecular weight of the atmosphere, $\mu$, directly affects the atmospheric scale height $H$ ($H \propto 1 / \mu$; \autoref{eq:atm_scale_height}), while the atmospheric mass primarily influences the base density, $\rhobase$. For a given planet and impact velocity, the values of $H$ and $\rhobase$ determine the size range of the impactors that most efficiently erode the atmosphere.
In the atmospheric erosion prescription of \citet{Shuvalov2009}, this relationship is encapsulated in the definition of the erosional efficiency factor, $\xi$ (\autoref{eq:erosional_efficiency_factor}), which determines the extent of atmospheric erosion (\autoref{eq:dm_atm_single}). For a given impact velocity and impactor mass, atmospheric mass loss peaks at $\xi\sim 350$ and falls for both higher and lower values of $\xi$, declining to less than $10\%$ of its peak value beyond the range $10 \lesssim \xi \lesssim 10^{5}$ (see Figure 6 of \cite{Shuvalov2009}). From \autoref{eq:erosional_efficiency_factor}, we find that $D^{3}  \propto \xi \matm / \mu^{2}$. So, as $\matm$ increases, the efficient erosion regime ($10 \lesssim \xi \lesssim 10^{5}$) shifts towards the larger impactor sizes, where we have more available mass in the debris, based on the adopted debris size distribution (\autoref{subsec:debris_evol}). If the efficient erosion regime remains below the maximum debris size ($\Dmax$), the atmospheric erosion per unit accreted mass increases (\autoref{eq:dm_atm}). However, a continued increase in $\matm$ may shift the efficient erosion regime to even larger impactor sizes exceeding $\Dmax$, leading to a decline in atmospheric erosion. This transition occurs at lower $\matm$ for lower values of $\mu$ (e.g., at  $\matm \simeq 0.7\ \matmearth$ for $\mu = 2.35$, while at $\matm \simeq 100\ \matmearth$ for $\mu = 29$) due to the $\mu$ dependence of $\xi$.

\autoref{fig:f_atm_m_atm_mu} shows the fraction of atmosphere remaining $1$ Gyr after the impact as a function of the initial atmospheric mass. For this analysis, we adopt $\fesc=0.01$ and $\Dmax = 100\ \km$, resulting in a total accreted mass of $0.0029\ \Mearth$. We consider two scenarios: one with an atmosphere having $\mu = 29$ similar to present-day Earth, and another with a primordial solar composition having $\mu = 2.35$. For the $\mu=29$ atmospheres, re-accretion of the debris causes complete atmospheric erosion within $\sim70$ Myr if the initial atmospheric mass, $\matmi \lesssim 5\ \matmearth$. For more massive atmospheres, the impacts from the debris are insufficient to erode the atmosphere entirely, given the total mass accreted. As expected, the fraction of the atmosphere retained increases with the initial atmospheric mass. In the case of the primordial atmosphere ($\mu=2.35$), atmospheric erosion is more efficient compared to the $\mu=29$ case for $\matmi \lesssim 5\ \matmearth$, and complete atmospheric erosion occurs relatively faster (see \autoref{fig:f_atm_m_atm_mu}). However, for more massive atmospheres, atmospheric erosion efficiency decreases when $\mu = 2.35$, whereas it increases for $\mu = 29$, as discussed in the previous paragraph. Hence, the fraction of atmosphere retained for $\mu = 2.35$ exceeds that of the atmosphere with $\mu = 29$.

These results indicate that the re-accretion of giant impact debris is highly efficient at eroding thin, terrestrial atmospheres, with the general trend remaining largely unaffected by variations in the mean molecular weight. In the context of a hypothetical Moon-forming collision with $\fesc = 0.01$ and $\Dmax=100\ \km$, our results show substantial atmospheric loss due to debris re-accretion, even if the post-impact Earth had an atmosphere over five times more massive than that of present-day Earth.

\subsection{Extending the Analysis to a Broader Parameter Space}\label{subsec:generalization}

In this section, we extend our analysis beyond the hypothesised giant impact on Earth and explore the general trends across a broader parameter space of $\Mp$, $\ap$ and $\matmi$. We consider a terrestrial planet with a bulk density of $\rhop = 5.5\ \gcc$, orbiting a Sun-like star ($\Mstar = \Msun$ and $\Lstar = \Lsun$), that undergoes a giant impact, ejecting $1\%$ of the planet's mass as debris ($\fesc = 0.01$) with a maximum size of $\Dmax=100\ \km$. Following the impact, the planet is assumed to retain an atmosphere with $\mu=29$ having a mass, $\matmi$.
\begin{figure}
\includegraphics[width=\columnwidth]{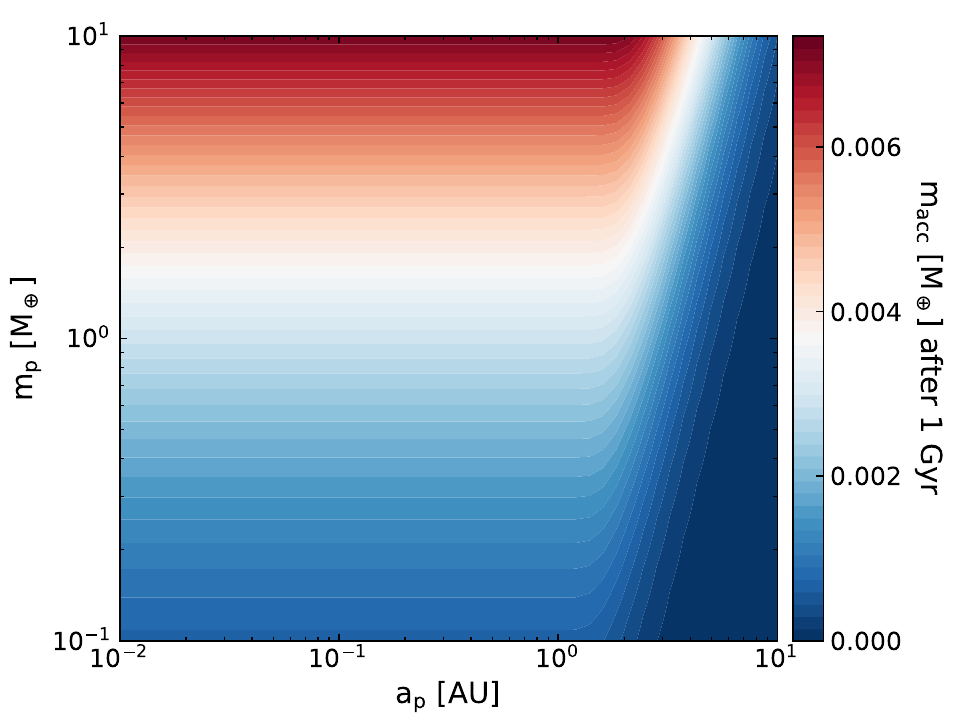}
\caption{Amount of debris re-accreted over $10^{9}$ years as a function of planet mass ($\Mp$) and semi-major axis ($\ap$). The planet is assumed to orbit a Sun-like star with an impact debris disk characterized by $\fesc=0.01$ with $\Dmax = 100\ \km$.}
\label{fig:atm_loss_ap_mp_macc}
\end{figure}
\begin{figure*}
\includegraphics[width=\textwidth]{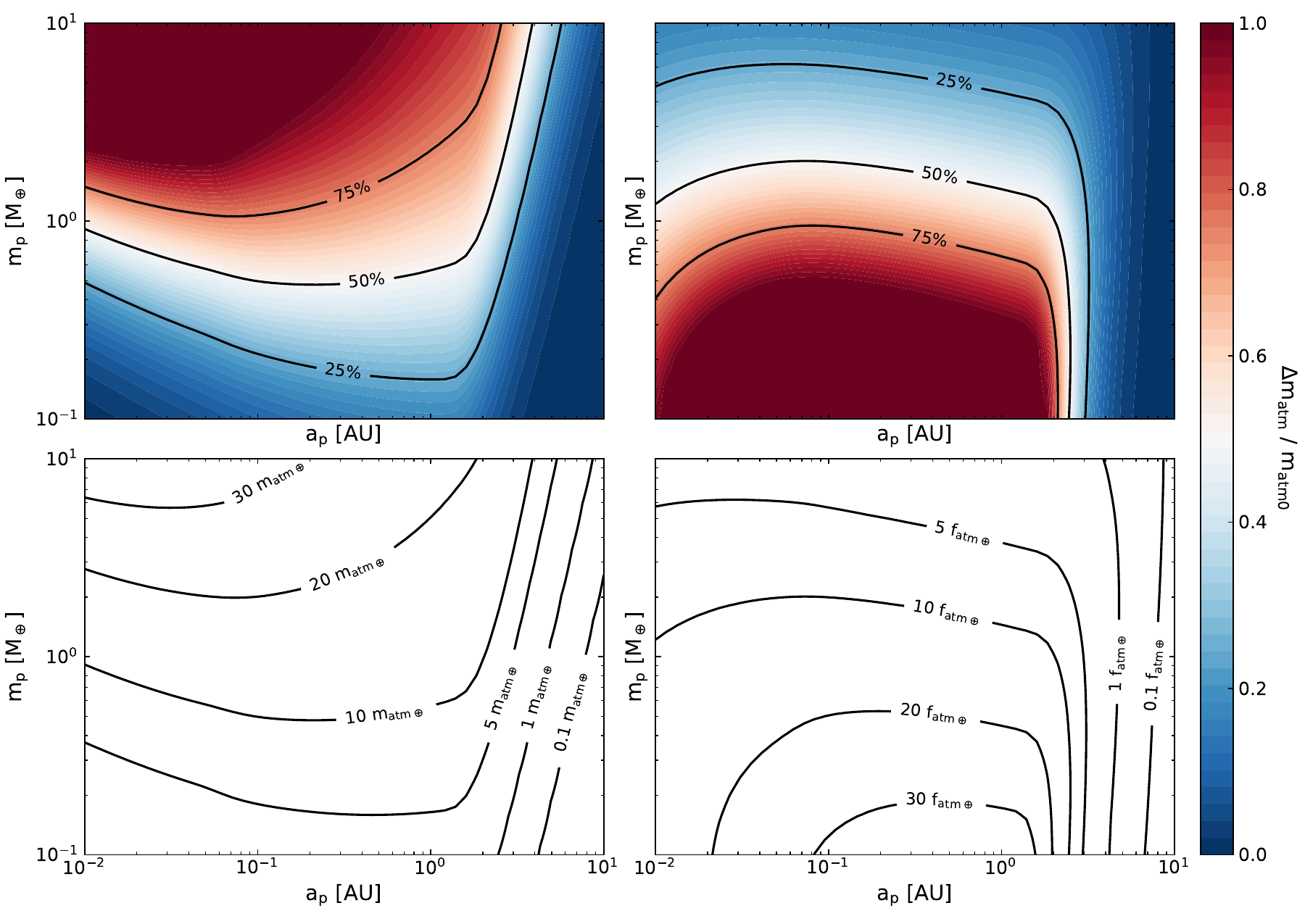}
\caption{
Post-giant-impact atmospheric erosion as a function of planet mass ($\Mp$) and semi-major axis ($\ap$) for varying initial atmospheric masses ($\matmi$). 
\textit{Top panels}: The colours represent the fraction of the initial atmospheric mass lost within $1$ Gyr following the primary giant impact for a planet with $\matmi = 10\ \matmearth$ (left) and for a planet with $\fatm = 10\ \fatmearth$ (right), with the black lines indicating $25\%$, $50\%$ and $75\%$ loss in each case. 
\textit{Bottom panels}: The lines represent $50\%$ atmospheric erosion contours for different initial atmospheric masses (see contour labels). Note that the spacing between contours representing different atmospheric erosion percentages (not explicitly shown) also varies with $\matmi$, typically exhibiting narrower spacing at lower $\matmi$ and wider spacing at higher $\matmi$. 
\textit{Left panels}: $\matmi$ is varied independent of $\Mp$. \textit{Right panels}: $\matmi$ scales with $\Mp$ as $\matmi = \fatm \Mp$.
All scenarios assume a Sun-like host star and an impact-generated debris disk with $\fesc=0.01$ with $\Dmax = 100\ \km$.
}
\label{fig:atm_loss_ap_mp_matm-4panel}
\end{figure*}

First, we consider the amount of debris re-accreted after $1$ Gyr and its dependence on the $\Mp - \ap$ parameter space (\autoref{fig:atm_loss_ap_mp_macc}). At times $t \gg \tauacc$, the debris disk reaches the end of its lifetime, with its mass depleted (\autoref{eq:mdisk_sol}), and re-accretion stops after accreting a total debris mass of $\macc \simeq \fesc \Mp \eta^{-1} \ln(1+\eta)$ (\autoref{eq:m_reacc}). Notably, $\eta$ is independent of orbital radius (as both $\tauacc$ and $\tauc$ scale as $\ap^4$) and has a weak dependence on the planet's mass (\autoref{eq:eta}). Consequently, the total debris mass accreted is also independent of the orbital radius of the planet and has only a weak dependence on the planet's mass, with more massive planets accreting more debris mass. Indeed, for our assumptions, all $0.1-10\ \Mearth$ planets undergoing giant impacts at $\ap \lesssim 1\ \au$ end up accreting a total debris mass with a factor of $2.4$ times more or $4.2$ times less than $0.003\ \Mearth$. However, at larger orbital radii ($\ap \gtrsim 2\ \au$), both the re-accretion and collisional depletion timescale become increasingly large, slowing down the evolution of the debris disk and its re-accretion rate. At such larger orbital distances, the disk's lifetime can exceed our chosen endpoint of $1$ Gyr. Consequently, the planet could potentially accrete more mass if its evolution continued longer. However, over such extended timescales, other physical processes (e.g., debris removal due to perturbation from neighbouring planets, atmospheric growth due to delivery of volatiles from asteroids or outgassing from the mantle) become increasingly significant and are outside the scope of this study. In what follows, it is worth bearing in mind that planets with orbital radii beyond $\sim 2\ \au$ have lower debris re-accretion (\autoref{fig:atm_loss_ap_mp_macc}) and remembering that this is because of the long timescales required for re-accretion which motivates our choice of simulation endpoint.

The planet's mass, orbital location, and atmospheric mass play important roles in the atmospheric erosion efficiency for a given size distribution of the impactors and the total mass accreted. These parameters affect the scale height of the atmosphere (\autoref{eq:atm_scale_height}), which in turn influences the size range (or $\xi$-range, \autoref{eq:erosional_efficiency_factor}) of the impactors that are most efficient in atmospheric erosion ($10 \lesssim \xi \lesssim 10^{5}$). From \autoref{eq:erosional_efficiency_factor}, we find $D^{3} \propto \xi \matm / (\ap \Mp^{4/3})$. So, as the orbital radius or the planet mass increases, or as the atmospheric mass decreases, the efficient erosion regime (corresponding to a fixed range of $10 \lesssim \xi \lesssim 10^{5}$) shifts towards smaller impactor sizes, and vice versa. Given the adopted size distribution of the debris (\autoref{subsec:debris_evol}), a greater proportion of the total debris mass is contained in the larger bodies. Hence, considering the parameter dependence discussed above, when the most efficient size range of the impactors is towards the larger debris but contained below the maximum debris size ($\Dmax$) in the disk, the atmospheric erosion per unit debris mass accreted increases. Conversely, the erosion efficiency diminishes as the most efficient size range shifts towards either smaller debris with a lower proportion of mass or larger debris sizes exceeding $\Dmax$ (also see the discussion in \autoref{ssubsec:atmos_variation}).

The top left panel of \autoref{fig:atm_loss_ap_mp_matm-4panel} shows the fraction of the initial atmosphere eroded due to the re-accretion of giant impact ejecta in the $\Mp-\ap$ parameter space for planets with initial atmospheric masses of $10\ \matmearth$. To understand the form of this plot, consider a $1\ \Mearth$ planet orbiting at different distances. This figure shows that the atmospheric erosion initially increases with orbital radius and then decreases. This is because the higher incident flux at lower orbital radii increases the atmospheric temperature and scale height. Consequently, at $\ap = 0.01\ \au$, the size distribution of the most erosionally efficient impactors peaks at $\sim 84\ \km$, with a significant proportion of the range extending beyond $\Dmax$ (up to $\sim 555\ \km$). For larger orbital radii, the range shifts to lower values (peaking at $\sim 39\ \km$ at $0.1\ \au$), coinciding with the debris size range holding a greater proportion of the mass. Therefore, atmospheric erosion increases per unit of accreted mass. However, for further increase in $\ap$, the efficient size range decreases towards the smaller debris size containing a lower proportion of the debris mass, reducing the atmospheric erosion per unit accreted mass. Since the total debris mass accreted is independent of the orbital radius for $\lesssim 2\ \au$, this trend directly translates to the total amount of atmosphere eroded. The sharp decrease in atmospheric loss beyond $\ap \gtrsim 2\ \au$ is due to the decline in total debris mass accreted within $1$ Gyr.

The dependence of atmospheric erosion on planetary mass is affected by both the atmospheric erosion efficiency and the amount of debris accreted. The dependence of atmospheric erosion efficiency on planet mass can be understood in a similar way to the dependence in $\ap$. Except in this case, as planetary mass increases, the total mass of accreted debris also increases (\autoref{fig:atm_loss_ap_mp_macc}), leading to more atmospheric erosion for massive planets for a given initial atmospheric mass. For example, at $\ap = 1\ \au$, a $0.5 \Mearth$ mass planet would lose $\sim 48\%$ of its atmosphere while a $2 \Mearth$ planet would lose $\sim 73\%$, assuming $\matmi = 10\ \matmearth$ in both cases (top left panel, \autoref{fig:atm_loss_ap_mp_matm-4panel}).

The atmospheric erosion in the case of initial atmospheric masses different from $10\ \matmearth$ can be understood in the same way. This is shown in the bottom left panel of \autoref{fig:atm_loss_ap_mp_matm-4panel} which shows the contours at which planets have $50\%$ of their atmosphere eroded due to debris re-accretion for the assumed initial atmospheric mass. Planets below these lines can be considered to retain a large fraction of their atmospheres despite re-accretion, whereas those above these lines suffer significant atmospheric erosion. As the initial atmospheric mass increases, the amount of debris required to erode the same fraction of the atmosphere correspondingly increases. Due to this, the $50\%$ atmospheric erosion contour lines (on the bottom left panel of \autoref{fig:atm_loss_ap_mp_matm-4panel}) of more massive atmospheres shift towards larger planetary masses, which is because they accrete more debris mass.

The right panels of \autoref{fig:atm_loss_ap_mp_matm-4panel} consider the case when the initial atmospheric mass scales with the planet's mass as $\matmi = \fatm \Mp$, where $\fatm$ represents a proportionality constant, treated as a parameter relative to Earth's atmospheric mass fraction $\fatmearth = \matmearth / \Mearth$. In this case the top right panel of \autoref{fig:atm_loss_ap_mp_matm-4panel}, representing $\fatm = 10\ \fatmearth$, shows that the fractional atmospheric erosion has a different dependence across the $\Mp-\ap$ parameter space to that of the $\matmi = 10\ \matmearth$ assumption of the top left plot. We find that atmospheric erosion decreases with planetary mass. For instance, at $\ap = 1\ \au$, a $0.5\ \Mearth$ mass planet with $\fatm = 10\ \fatmearth$ would lose $\sim 85\%$ of its atmosphere, whereas a $2\ \Mearth$ planet with the same $\fatm$ would lose $\sim 41\%$. This trend can be attributed to the fact that for less massive planets, the re-accretion timescale decreases relative to the collisional depletion timescale (i.e., $\eta$ decreases, \autoref{eq:eta}). Consequently, the planets can accrete a greater fraction of the proportionally less massive debris disk (\autoref{eq:m_reacc}), whereas the atmospheric mass subject to erosion scales linearly with the planet's mass, resulting in more debris accretion per unit atmospheric mass for less massive planets. Considering the bottom right panel of \autoref{fig:atm_loss_ap_mp_matm-4panel}, we find that as $\fatm$ increases, the required amount of debris correspondingly increases, and so the $50\%$ atmospheric erosion contour lines shift towards lower planetary masses, which accrete larger $\macc / \Mp$.

Overall, our results indicate that the planets growing via giant impacts within $2\ \au$ are likely to experience significant atmospheric erosion due to the re-accretion of the giant impact ejecta. We find that planets more massive than $0.4\ \Mearth$ are likely to lose the majority of their atmospheres if they start with $\lesssim 5\ \matmearth$, with more massive planets losing more (bottom left panel, \autoref{fig:atm_loss_ap_mp_matm-4panel}). Conversely, if the initial atmospheric mass scales with the planet's mass, sub-Earth-mass planets with $\fatm \sim 10 \fatmearth$ are likely to lose the majority of their atmosphere, with lower-mass planets losing the majority of their atmosphere for even larger $\fatm$ (bottom right panel, \autoref{fig:atm_loss_ap_mp_matm-4panel}).

\section{Discussion}\label{sec:discussion}

Our results show that the re-accretion of the giant impact ejecta can drive significant atmospheric erosion on a post-giant-impact planet (\autoref{sec:res}). In this section, we discuss the importance of the velocity dispersions in the debris disk (\autoref{subsec:discuss:v_rel}), the influence of the neighbouring planets (\autoref{subsec:discuss:other_planets}), processes that replenish the atmosphere (\autoref{subsec:discuss:replenishment_atmos}), and the geochemical implications of re-accretion of impact-generated debris (\autoref{subsec:discuss:geo_chem}).

\subsection{Relative Velocity Distribution in the Debris Disk}\label{subsec:discuss:v_rel}

The velocity distribution of the debris, $\vrel$, plays a crucial role in the collisional evolution of the disk and atmospheric erosion. A larger velocity dispersion between the debris makes the collision rate higher, thereby reducing the collisional lifetime of the debris disk (see \aref{app:tau_cc}) and reducing the total mass accreted by the planet. However, a larger velocity dispersion also results in higher impact velocities of debris onto the planet, increasing the atmospheric erosion in each impact.

In this study, we consider the relative velocities of debris based on numerical simulations of the canonical Moon-forming impact. Specifically, we adopt the empirical velocity distribution fit of the debris distribution reported by \citet{Jackson_2014}, derived from the smoothed particle hydrodynamics (SPH) simulations of \citet{2009_Marcus}. Although the limited resolution of these early simulations (the largest debris constituted by only a handful of particles) warrants high-resolution simulations (a task beyond the scope of this study), we do not expect the overall velocity distribution to differ substantially with increased resolution. \citet{Jackson_2014} suggest that the ejecta from the canonical Moon-forming impact follows a Gaussian velocity distribution with a mean of zero and a standard deviation of $\sigv \simeq 0.46 \vesc$ in the collision frame.
We use the dispersion of the debris velocities to estimate $\vrel$, resulting in values lower than those traditionally adopted. For instance, \cite{Genda_2015} adopted $\vrel \sim \vesc$, while \citet{Wyatt_2007, Wyatt_2017} adopted $\vrel \sim 1.5 \vesc$. We argue that our $\vrel$ is more accurate, since it is based on the velocity distribution obtained in a giant impact simulation, whereas the other estimates were order of magnitude estimations with order unity uncertainties.

Given the strong dependence on the relative velocities (\autoref{eq:tau_c_basic}), our collisional depletion timescale is larger by $\sim23$ times compared to \citet{Wyatt_2017}, solely due to lower $\vrel$. On the other hand, by accounting for the cratering collisions, our calculation has also reduced $\tauc$ by a factor of $3.3$ (see \aref{app:tau_cc}), resulting in a net $\tauc$ that is  $\sim 7$ times longer than the estimates of \cite{Wyatt_2007, Wyatt_2017}. In contrast, due to the gravitational focusing term in \autoref{eq:v_imp}, the reduction in $\vrel$ does not result in as large a change in $\vimp$, which only decreases by $\sim 40 \%$. Altogether, the lower estimate of $\vrel$ results in a significant increase in the re-accretion of the debris, leading to substantial atmospheric erosion.

In the case of more energetic giant impacts with velocities much higher than those typically assumed during the canonical Moon-forming impact, the velocity dispersion of the debris would likely be higher, shortening the disk lifetime. However, high-energy collisions are also likely to eject more debris, subject to re-accretion at higher impact velocities, potentially counterbalancing the effects of lower disk lifetime on atmospheric erosion. The extent of atmospheric erosion in these scenarios would depend on the precise values of the parameters $\vrel$ and $\fesc$, which can only be determined from detailed numerical simulations.

\subsection{Effect of/on Neighbouring Planets}\label{subsec:discuss:other_planets}

\begin{figure}
\includegraphics[width=\columnwidth]{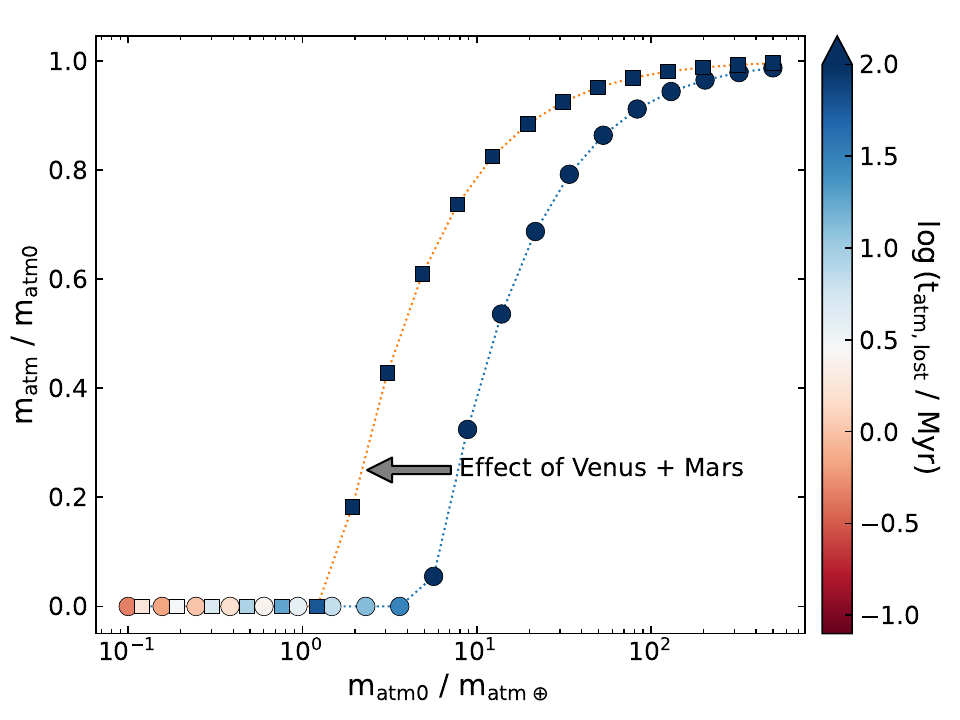}
\caption{Fraction of atmosphere retained $1$ Gyr after a hypothetical Moon-forming impact with $\fesc = 0.01$ and $\Dmax=100\ \km$, as a function of the initial atmospheric mass ($\matmi$), analogous to \autoref{fig:f_atm_m_atm_mu}. Filled circles represent the scenario where the Earth is the only accreting planet. The filled squares represent the scenario where the mass accreted by Earth is reduced by a factor of $3$, accounting for the influences of neighboring planets. In both cases, we assume $\fesc = 0.01$, $\Dmax=100\ \km$, and $\mu = 29$.}
\label{fig:f_atm_m_atm_vemj}
\end{figure}

The analysis presented in this study considers a single planet interacting with the debris. However, the presence of any neighbouring planets may have significant implications for the evolution of the debris and its re-accretion onto the parent planet. The effects of nearby planets in the context of the canonical Moon-forming collision were numerically studied by \cite{Jackson_Wyatt_2012}. They found that planets far form Earth, such as Mercury or those beyond Jupiter, have negligible influence on the debris. Jupiter, having a large interaction cross section due to its enormous mass, imparts a strong kick on the interacting debris, throwing them into the inner Solar system or ejecting them entirely. Such interactions truncate the disk interior to Jupiter's orbit, preventing their direct interaction with the outer planets. In the simulations conducted in \cite{Jackson_Wyatt_2012} $\sim 8\%$ of the debris was ejected due to encounters with Jupiter within $10$ Myr. However, these simulations did not take into account the depletion of the debris mass due to collisional grinding, which would result in an even smaller fraction of the initial debris mass being ejected.

Immediate neighbours such as Venus and Mars have a more significant influence on the debris. Venus, at close orbital separation, will spread the debris inwards and accrete a substantial fraction itself, comparable to the amount accreted by Earth \citep{Jackson_Wyatt_2012, Gladman_1996}. Mars, because of its lower mass and relatively larger separation from Earth, has a smaller effect on the debris population, accreting less than $1\%$ of the total debris. Overall, \citet{Jackson_Wyatt_2012} found that the re-accretion of debris particles onto the Earth decreased by almost a factor of 3 in the presence of the neighbouring planets.

To emulate the influence of these neighbouring planets in our analysis, we consider a simplified model where the debris mass re-accreted by Earth is reduced by a factor of 3 and evaluate its consequence on atmospheric erosion
\footnote{Note that the factor of 3 reduction in accreted particles found by \citet{Jackson_Wyatt_2012} may not directly correspond to a proportional decrease in the total mass accreted because the perturbations from neighbouring planets affect the debris at the late stages of evolution, at which point a substantial portion of the debris mass has already been depleted due to collisional grinding. Therefore, this factor of three reduction in mass accretion should be treated as a conservative lower bound.}
. The resulting atmospheric erosion is shown in Figure \ref{fig:f_atm_m_atm_vemj} as a function of initial atmospheric mass considering $\fesc = 0.01$, $\Dmax=100\ \km$, and $\mu=29$. As expected, a reduced accretion leads to reduced atmospheric losses, particularly for planets with substantial initial atmospheres. In this scenario, Earth experiences $\gtrsim 20\%$ atmospheric erosion for $\matmi \lesssim 10\ \matmearth$, and complete atmospheric erosion occurs for $\matmi \lesssim \matmearth$ on a relatively longer timescale ($\sim 30$ Myr). Complete erosion of more massive atmospheres is likely for larger $\fesc$ or $\Dmax$. 
Since Venus is also expected to accrete comparable amounts of debris as Earth \citep{Jackson_Wyatt_2012}, we use our model to estimate the effects of this accretion on the Venusian atmosphere. Given that Venus has a significantly more massive atmosphere ($\matmvenus / \matmearth \simeq 94$), it loses only $2.6\%$ of its atmosphere due to debris accretion. However, the accretion of such large amounts of debris would result in the deposition of a several-km-thick layer of material with Earth-Moon origin.
On the other hand, although Mars is estimated to accrete only $1.45\times10^{-5}\ \Mearth$, approximately $1 / \ 67$th of debris mass accreted by Earth \citep{Jackson_Wyatt_2012}, this is still three orders of magnitude larger than its present-day atmosphere. As a result, even this limited debris accretion would have been sufficient to entirely erode Mars' present-day atmosphere and potentially remove up to three times its current atmospheric mass. This implies that, if it released substantial non-vaporised debris, Mars must have obtained its atmosphere since the Moon-forming impact or started with a much more massive atmosphere \citep[e.g.,][]{2025_Joiret}. However, the observed Xenon depletion and isotope fractionation in the modern Martian atmosphere support a impact driven atmospheric evolution scenario, favouring the former possibility \citep[][]{Shorttle_2024}.

Conversely, debris produced by giant impacts in neighbouring planets can also be accreted by the Earth, resulting in additional atmospheric erosion. For instance, \citet{Gladman_1996} found that $7.5\%$ of meteorites originated at Mars may impact Earth. Therefore, from a giant impact event on Mars that ejected $1\%$ of its mass in solid debris with $\Dmax=100\ \km$, Earth would accrete $8 \times 10^{-5} \Mearth$ of Martian material, neglecting collisional depletion of the debris
\footnote{The collisional depletion timescale for Mars under the single planet assumption of \autoref{eq:tau_c} is $\sim 495$ Myr. Additionally, perturbations from neighbouring planets would disperse the disk, thereby reducing the collision rate and further extending the depletion timescale. However, the Martian meteorites take a much shorter time to reach Earth as evidenced by their cosmic ray exposure age \citep[$\lesssim 15$ Myr,][]{Gladman_1996}. Consequently, collisional depletion is expected to have a negligible effect on Martian debris reaching Earth.}.
This level of accretion would erode $13\%$ of Earth's present-day atmosphere and cover the Earth's surface in a $\sim 170\ $m deep layer. If Venus suffered a giant impact ejecting $1\%$ of its mass in solid debris, and if Earth's influence on the resulting debris were analogous to Venus's influence on the debris from Earth, both Earth and Venus would accrete $0.0009 \Mearth$ of debris. The re-accretion of this debris would lead to the erosion of only $2.4\%$ of Venus's dense atmosphere, whereas Earth's entire atmosphere would be lost. Hence, the presence of Earth's atmosphere sets constraints about such a late giant impact on Venus.

While known exo-planetary systems typically have very diverse orbital architectures compared to the Solar System, the neighbouring planets of the progenitor (if present) are similarly expected to exert a strong influence on the debris dynamics \citep[e.g.,][]{Wyatt_2016}. Nonetheless, the general trends observed for the Solar system will continue to hold with revised timescales according to the orbital radius.

\subsection{Replenishment of the Atmosphere} \label{subsec:discuss:replenishment_atmos}

\begin{figure}
\includegraphics[width=\columnwidth]{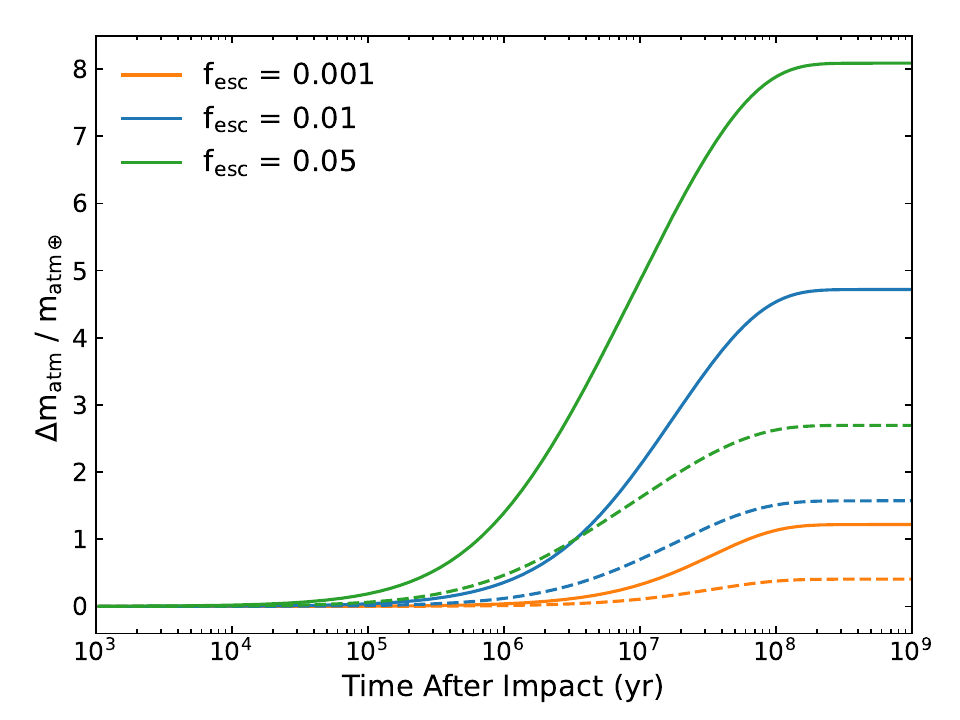}
\caption{Atmospheric mass loss over time due to impacts from debris, assuming instant replenishment of the atmosphere. This gives us an estimate of how much volatile can be potentially lost from the total volatile budget of the Earth due to re-accretion of giant impact debris. Corresponding dashed lines in the same colours represent the scenario where debris re-accretion is reduced by a factor of 3, consistent with the reduced accretion in the presence of neighbouring planets \citep{Jackson_Wyatt_2012}. In both cases, we assume $\Dmax=100\ \km$ and $\mu = 29$.}
\label{fig:evol_fixed_atm_fesc}
\end{figure}

In this study, we have demonstrated that the re-accretion of the giant impact ejecta is a significant mechanism contributing to atmospheric erosion in post-giant-impact planets. However, our simplified model does not account for various physical processes that can potentially contribute to atmospheric growth. For example, the impactors considered in our study originated from the mass ejected during the violent giant impact events which were assumed to have little to no volatile content. However, Earth has also experienced significant impacts from the asteroids and comets \citep[e.g., during late veneer;][]{Wang_2013, Morbidelli_2018}, which are likely to deliver additional volatiles and drive atmospheric growth. 
Moreover, isotopic signatures of atmospheric noble gases, particularly xenon (Xe), suggest a cometary contribution \citep[e.g.,][]{Marty_2017}. The timing of cometary volatile delivery remains highly uncertain, but likely overlaps with the phase of debris re-accretion following the Moon-forming impact \citep[e.g.,][]{Joiret_2023}.
The coupled effects of volatile delivery and atmospheric erosion have been studied by \citet{Wyatt_2020} and \citet{Sinclair_2020},  who found that, over a wide range of initial atmospheric masses and compositions, the resulting atmosphere tends to converge toward a mass and composition similar to that of present-day Earth. Another potentially significant source of a secondary atmosphere may originate from the impact-triggered outgassing of the volatiles stored in the mantle \citep[e.g.,][]{Elkins-Tanton_2008, Lebrun_2013, Marchi_2016, Schlichting_Mukhopadhyay_2018}. The substantial energy imparted by the impacts would partially melt the planet's surface, if not already molten during previous impacts or the giant impact event, releasing volatiles from the local magma ocean and enabling their loss through atmospheric erosion. Earth's geological records provide evidence of multiple such phases of mantle degassing and atmospheric erosion, contributing to the fractionation of specific isotope ratios within Earth's mantle \citep[e.g.,][]{Tucker_Mukhopadhyay_2014, Schlichting_Mukhopadhyay_2018}. In certain circumstances, the volatiles released from outgassing can not only compensate for, but potentially exceed, the impact-driven atmospheric losses \citep[e.g.,][]{Schlichting_2017, Sinclair_2020}.

Given the significance of such effects, we investigate the potential loss from the planet's volatile inventory due to the re-accretion of the giant-impact ejecta. We consider the simplified case where the atmosphere eroded due to debris re-accretion is replenished within a short timescale, such that the total atmospheric mass remains unchanged over time. \autoref{fig:evol_fixed_atm_fesc} shows the atmospheric erosion for this model of the Moon-forming event, assuming $\Dmax = 100\ \km$ and $\matmi = \matmearth$. Under such conditions, Earth could lose a total of $\sim 4.7\ \matmearth$ of volatiles for $\fesc = 0.01$, which increases to $\sim 8\ \matmearth$ for $\fesc = 0.05$. A more massive initial atmosphere would result in even higher volatile losses due to increased erosion efficiency; for instance, a scenario with $\matmi = 10\ \matmearth$ results in $\sim 6.4\ \matmearth$ lost when $\fesc = 0.01$. Even when considering a reduced debris re-accretion due to the influence of neighbouring planets, volatile losses remain significant. In this simplified model, we do not consider the scenario where atmospheric growth from various processes has a different rate relative to the erosion, nor follow the change in atmospheric composition and properties over time, which may affect the atmospheric erosion rate itself. We leave the exploration of atmospheric evolution accounting for all these competing processes for a future study.

\subsection{Geochemical Implications} \label{subsec:discuss:geo_chem}

In this study, we have shown that a significant fraction of ejected material from giant impacts can eventually be re-accreted onto planets. This re-accretion of giant impact debris has profound implications not only for the atmospheric evolution of terrestrial planets, as shown in this study, but also for their geochemical evolution.

Geochemical evidence indicates that Earth's volatile element inventories were shaped by a balance between the initial abundances in accreting bodies and their losses \citep[e.g.,][]{Mukhopadhyay_Parai_2019}. Several geochemical signatures point to substantial atmospheric loss during the giant impact phase of Earth's accretion. For instance, the depletion of radiogenic $^{129}$Xe produced from the decay of now-extinct $^{129}$I (half-life of $15.7$ Myr), isotopic fractionation of helium and neon isotopes (particularly $^{3}$He/$^{22}$Ne and $^{20}$Ne/$^{22}$Ne), and anomalous elemental ratios such as H/N and potentially F/Cl, collectively suggest extensive volatile loss during this period \citep[e.g.,][]{Tucker_Mukhopadhyay_2014, Schlichting_Mukhopadhyay_2018, Mukhopadhyay_Parai_2019}. However, giant impacts alone would generally be inefficient at removing atmospheres \citep[e.g.,][]{Schlichting_2015}. Therefore, additional mechanisms causing atmospheric erosion, such as the re-accretion of impact-generated debris \citep[which was likely abundant during this period;][]{Genda_2015}, could assist in explaining the volatile depletions and fractionations observed in geochemical records.

It has also been suggested that a significant portion of the Earth's water was accreted before the Moon-forming impact \citep[e.g.,][]{Halliday_2013, Greenwood_2018}. In such a scenario, the Moon-forming giant impact likely vaporised the pre-existing water, producing a briefly sustained massive steam atmosphere \citep[$\sim30-100$ bar;][]{Zahnle_1988}. Our results indicate that the atmospheric erosion due to debris re-accretion would not be efficient enough to remove a large fraction of such a substantial atmosphere (see \autoref{fig:f_atm_m_atm_vemj}). Only after condensation of the post-impact steam atmosphere would Earth have a sufficiently low mass atmosphere ($\lesssim 5$ bar) that could be substantially eroded by re-accretion. Thus, if water delivery predated the impact, the atmospheric erosion process modelled here would have minimal effect on its overall retention. This is in contrast to volatile elements like noble gases, the incondensable nature of which will leave them in the atmosphere after Earth's oceans have reformed. These hyper-volatiles are the elements most prone to removal by re-accretion - a phenomenon, which coupled with later delivery, could help explain the discrepancy in noble gas isotope composition between Earth's atmosphere and interior \citep[e.g.,][]{Halliday_2013}.

Another key archive of late accretion is provided by Earth's HSE budget. Impactors containing metallic iron could be oxidized upon re-accretion due to reactions with surface water or a relatively oxidized mantle. This oxidation prevents the iron from reaching the core and instead results in its retention within the mantle, contributing to the abundance of highly siderophile elements (HSEs) on Earth's mantle \citep{Genda_2017, Genda_Brasser_Mojsis_2017}. Small fragments (mm-sized), in particular, are susceptible to such reactions, while larger impactors are more likely to penetrate the mantle and ultimately segregate their metal to Earth's core. The reaction between metallic iron and water also produces H$_{2}$ that has significant implications for the origin of life on early Hadean Earth \citep[e.g.,][]{Urey_1952, Miller_1953, Zahnle_2020, Itcovitz_2022}. \citet{Genda_Brasser_Mojsis_2017} attribute the formation of such small metallic fragments to the disintegration of the impactor core during oblique giant impacts. We postulate that such small metallic fragments can also be produced via collisional grinding of larger debris containing metallic iron. Giant impact events may also produce a substantial amount of unbound mass in vapour, which will condense to mm-cm-sized droplets \citep[e.g.,][]{Melosh_Vickery_1991, Johnson_Melosh_2014}. Although such particles are incapable of eroding the atmosphere and were excluded from our analysis, they are likely to be efficiently oxidized upon re-accretion, contributing to both the mantle’s HSE inventory and the redox state of the planet.

Given the solar system terrestrial planets likely formed via series of giant impacts \citep[e.g.,][]{2016_Quintana, Raymond_Morbidelli_2022, Clement_2024}, repeated episodes of re-accretion could modulate the redox state of Earth over time. Fully accounting for the origin, size distribution and composition, consequences for Earth of its re-accreted small debris are beyond the scope of the paper. However, this cumulative effect may have played a critical role in the establishment of habitable conditions, particularly in the presence of early oceans that could act as oxidizing agents.


\section{Conclusions}\label{sec:conclusions}

The evolution of debris ejected during giant impacts is a critical yet under-explored aspect of the final stages of planet formation. Giant impacts are thought to be the primary mechanism of growth during the final stage of terrestrial planet formation and have profound implications for planetary composition, structure, and orbital properties. While advanced numerical simulations have successfully quantified immediate post-impact outcomes, such as mass loss and atmospheric erosion, the prohibitive computational cost of these simulations has limited the investigation of long-term post-impact effects. In this study, we attempted to address this gap by investigating the long-term evolution of giant impact debris and its implications for the progenitor.
Our results show that a significant fraction of the non-vaporised debris launched by giant impacts is likely to be re-accreted to the progenitor, largely independent of the specifics of the collision or assumptions about the debris properties. Depending on the total mass and sizes of the non-vaporised debris, the secondary impacts during re-accretion can significantly erode the planet's atmosphere, and thereby reduce the planet's volatile content.

Re-accretion of giant impact debris is a process that may have significant implications for the Earth. For instance, under plausible assumptions, the re-accretion of debris from the canonical Moon-forming collision could erode an atmosphere similar to that of present-day Earth over tens of millions of years. In general, any planet growing via giant impacts within $2\ \au$ is likely to experience significant post-impact atmospheric erosion unless the initial atmosphere was at least $5$ times more massive than Earth's, or the ejecta produced by such impacts are either small ($\ll10\ \km$) or predominantly vaporized.
These results highlight the importance of secondary effects of giant impacts, which extend far beyond the immediate aftermath of these events. Our study shows that giant impacts can be more effective at eroding atmospheres than previously thought. The impacts during the re-accretion of debris are an important mechanism of atmospheric erosion, and are crucial for investigations of long-term atmospheric evolution of Earth-like planets.


\section*{Acknowledgements}

We thank the anonymous referee for insightful comments and constructive suggestions. T.G. gratefully acknowledges support from the Leverhulme Centre for Life in the Universe at the University of Cambridge through the Joint Collaborations Research Project Grant G112026, LGAG/393.

\section*{Data Availability}

 The data underlying this article will be shared on reasonable request to the corresponding author.


\bibliographystyle{mnras}
\bibliography{ref}



\appendix
\section{Collisional Depletion Timescale of Debris}\label{app:tau_cc}

The number of bodies with size between $D$ and $D + dD$ is $n(D)dD$, and we assume the size distribution follows a power law:
\begin{equation}
    n(D) = KD^{-\alpha},
\end{equation}
with $\alpha = 3.5$ for steady state collisional cascade \citep{Wyatt_2007, Wyatt_2008}, and $K$ being a scaling parameter. Assuming the disk mass, $\Mdisk$, is primarily contributed by the largest bodies ($\alpha < 4$) that are significantly larger than the smallest ones, we obtain
\begin{equation}
    K \simeq 6 (4 - \alpha) \frac{\Mdisk}{\pi \rho}\Dmax^{\alpha - 4},
\end{equation}
where $\rho$ is the density and $\Dmax$ is the diameter of the largest bodies.

We adopt the collisional fragmentation model described in \citet{Wyatt_2002}. According to this model, the mass fraction of the largest remaining fragment is given by
\begin{equation}
\begin{aligned}
    \flr & = 1 - 0.5 \phi\ \ \text{for cratering collisions ($\phi < 1$),}  \\
         & = 0.5 \phi^{-\beta}\ \ \ \ \text{for catastrophic collisions ($\phi \geq 1$),}
\end{aligned}
\end{equation}
where $\beta = 1.24$, and $\phi$ is the ratio of the specific incident energy, $Q$, to the dispersal threshold, $\Qdstar$. For $D \gg \Dim$,
\begin{equation}\label{eq.phi_Xc}
    \phi(D,\Dim) = \frac{\vrel^{2}}{2 \Qdstar} \left(\frac{\Dim}{D}\right)^{3} = \left(\frac{\Dim}{\Xc D}\right)^{3}
\end{equation}
where $\Xc D$ is the minimum diameter of an impactor required to cause a catastrophic collision (i.e., $\phi(D, \Xc D) =1$).

The rate of collisions experienced by an object with size $D$ from impactors in the size range $\Dim$ to $\Dim + d\Dim$ is 
%
\begin{align}
    \Rcol(D, \Dim)d\Dim & = A \Dim^{2-\alpha} ( 1 + D/\Dim)^{2} d\Dim,
\end{align}
where,
\begin{equation}
    A = \frac{K \pi}{4 V} \vrel = \frac{3(4-\alpha)}{2 \rho V} \Mdisk \Dmax^{\alpha-4} \vrel,
\end{equation}
$\vrel$ is the relative velocity and $V$ denotes the volume of the disk. Here, we have ignored the gravitational focussing effects, as the relative velocities are assumed to be high. Under these conditions, the coalescence of bodies is also negligible, and the rate of mass loss from the objects of size $D$, having total mass $\Mtot(D)$, due to collisions with objects of size between $\Dim$ and $\Dim + d\Dim$ is estimated as:
\begin{equation}
\begin{aligned}
    \Rml(D, \Dim)d\Dim & = \Rcol(D, \Dim)(1 - \flr(D, \Dim)) \Mtot(D) d\Dim. \\
\end{aligned}
\end{equation}
Therefore, the total mass loss rate for objects of size $D$, due to collisions with impactors from $\Dbl$, the radiation-pressure blowout size, up to $\Dmax$, is
\begin{align}\label{eq:mass_loss_rate}
    \Rml(D) & = \int_{\Dbl}^{\Xc D} \Rml(D, \Dim) \,d\Dim + \int_{\Xc D}^{\Dmax} \Rml(D, \Dim) \,d\Dim \nonumber\\
            & = \Rmlcr(D) + \Rmlcc(D) .
\end{align}
The first term represents the rate of the cratering collisions, while the second term gives the catastrophic collision rate. Now the cratering collision rate,
\begin{align}\label{eq:cratering_rate}
    \Rmlcr(D) & = \int_{\Dbl}^{\Xc D} \Rcol(D, \Dim) 0.5 \phi(D, \Dim) \Mtot(D) \,d\Dim \nonumber\\
              & = \frac{0.5 A \Mtot(D)}{(\Xc D)^{3}} \left[\frac{\Dim^{6-\alpha}}{6-\alpha} + \frac{2 D \Dim^{5-\alpha}}{5-\alpha} + \frac{D^2 \Dim^{4-\alpha}}{4-\alpha}\right]_{\Dbl}^{\Xc D} .
\end{align}
For $\alpha < 4$, the upper limit dominates:
\begin{equation}
    \Rmlcr(D) \simeq 0.5 A \Mtot(D) D^{3-\alpha} \left[\frac{\Xc^{3-\alpha}}{6-\alpha} + \frac{2 \Xc^{2-\alpha}}{5-\alpha} + \frac{\Xc^{1-\alpha}}{4-\alpha}\right].
\end{equation}
Further since $\alpha > 3$ and $\Xc \ll 1$, the last term will be the most significant:
\begin{equation}\label{eq:cratering_rate_approx}
    \Rmlcr(D) \simeq \frac{A \Mtot(D)}{2(4-\alpha)} D^{3-\alpha}\Xc^{1-\alpha}.
\end{equation}
The rate of catastrophic collisions (from \autoref{eq:mass_loss_rate}),
\begin{align}\label{eq:catastrophic_rate}
    \Rmlcc(D) & = \int_{\Xc D}^{\Dmax} \Rcol(D, \Dim) (1 - 0.5 \phi^{-\beta}(D, \Dim)) \Mtot(D) \,d\Dim \nonumber\\
              & = \Rmlccone(D) + \Rmlcctwo(D).
\end{align}
Now the first term,
\begin{align}\label{eq:catastrophic_rate_1}
    \Rmlccone(D) & =  \int_{\Xc D}^{\Dmax} \Rcol(D, \Dim) \Mtot(D) \,d\Dim \nonumber\\
    & = A \Mtot(D) \left[\frac{\Dim^{3-\alpha}}{3-\alpha} + \frac{2 D \Dim^{2-\alpha}}{2-\alpha} + \frac{D^2 \Dim^{1-\alpha}}{1-\alpha}\right]_{\Xc D}^{\Dmax}.
\end{align}
For $\alpha > 3$, the lower limit dominates:
\begin{equation}
    \Rmlccone(D) \simeq - A \Mtot(D) D^{3-\alpha} \left[\frac{\Xc^{3-\alpha}}{3-\alpha} + \frac{2 \Xc^{2-\alpha}}{2-\alpha} + \frac{\Xc^{1-\alpha}}{1-\alpha}\right].
\end{equation}
Further since $\alpha > 3$ and $\Xc \ll 1$, the last term will be dominant:
\begin{equation}\label{eq:catastrophic_rate_1_approx}
    \Rmlccone(D) \simeq \frac{A \Mtot(D)}{\alpha-1} D^{3-\alpha}\Xc^{1-\alpha} .
\end{equation}
The second term from \autoref{eq:catastrophic_rate},
\begin{align}\label{eq:catastrophic_rate_2}
    \Rmlcctwo(D) & = - 0.5 \int_{\Xc D}^{\Dmax} \Rcol(D, \Dim) \phi^{-\beta}(D, \Dim) \Mtot(D) \,d\Dim \nonumber\\
    & = - 0.5 A \Mtot(D) (\Xc D)^{3 \beta} \nonumber\\
    & \times \left[\frac{\Dim^{3-\alpha-3\beta}}{3-\alpha-3\beta} + \frac{2 D \Dim^{2-\alpha-3\beta}}{2-\alpha-3\beta} + \frac{D^2 \Dim^{1-\alpha-3\beta}}{1-\alpha-3\beta}\right]_{\Xc D}^{\Dmax} .
\end{align}
Again for $\alpha > 3$ and $\Xc \ll 1$, the lower limit will dominate and the last term will be most significant:
\begin{equation}\label{eq:catastrophic_rate_2_approx}
    \Rmlcctwo(D) \simeq  - \frac{A \Mtot(D)}{2(\alpha+3\beta-1)} D^{3-\alpha}\Xc^{1-\alpha} .
\end{equation}
Now using \autoref{eq:catastrophic_rate_1_approx} and \autoref{eq:catastrophic_rate_2_approx}, we get from \autoref{eq:catastrophic_rate},
\begin{align}\label{eq:catastrophic_rate_approx}
    \Rmlcc(D) & = \Rmlccone(D) + \Rmlcctwo(D) \nonumber\\
              & \simeq \Rmlccone(D) \left[1 - \frac{\alpha-1}{2(\alpha+3\beta-1)}\right].
\end{align}
Using \autoref{eq:mass_loss_rate}, \autoref{eq:cratering_rate_approx}, and \autoref{eq:catastrophic_rate_approx}, we finally get
\begin{align}
    \Rml(D) & = \Rmlcr(D) + \Rmlcc(D) \nonumber\\
              & \simeq \Rmlccone(D) \left[\frac{\alpha -1}{2(4-\alpha)} +  1 - \frac{\alpha-1}{2(\alpha+3\beta-1)}\right].
\end{align}
Using $\beta = 1.24$ and the steady state value of $\alpha = 3.5$, we get,
\begin{equation}\label{eq:mass_loss_rate_final}
    \Rml(D) = 3.3\ \Rmlccone(D).
\end{equation}
Now assuming the disk mass is initially dominated by the largest planetesimals ($\Mdisk \simeq \Mtot(\Dmax)$), we estimate the collisional depletion timescale from the mass loss rate of the largest bodies, given by,
\begin{align}\label{eq:tau_c_basic}
    \tauc & \simeq \frac{\Mdisk}{\Rml(\Dmax)} = \frac{1}{3.3} \frac{2(\alpha -1)\rho V}{3(4-\alpha)} \Mdisk^{-1} \Dmax \Xc^{\alpha-1} \vrel^{-1}.
\end{align}
Note that, under identical assumptions about debris disk and the relative velocities of the bodies, the collisional depletion timescale derived in \autoref{eq:tau_c_basic} is $3.3$ times shorter than the estimate reported in \cite{Wyatt_2007}, where collisional depletion due to cratering collisions was not taken into account. However, this reduction factor is shorter than the estimates of \citet{Kobayashi_2010}, who estimated $5-7$ times reduction compared to \citet{Wyatt_2007}, when cratering collisions are included. This discrepancy likely originates from the differences in the fragmentation models adopted.

In this study, we estimate the relative velocity of the bodies from the velocity dispersion in the debris disk as $\vrel \sim \sigv \approx 0.46 \vesc$, consistent with the results from the numerical simulations of the canonical Moon-forming impact  \citep{Jackson_Wyatt_2012, Jackson_2014, Wyatt_2016}. Assuming a toroidal geometry for the disk, with its width determined by the planet's escape velocity, and recalling $\Xc = (2\Qdstar/\vrel^2)^{1/3}$ (\autoref{eq.phi_Xc}, with $\phi =1$), we obtain,
\begin{align}\label{eq:tauc_final_form}
    \tauc \simeq \frac{243}{G^{4/3}} \ap^{4} \Mp^{-2/9} \rhop^{-1/9} \rho \Dmax \Qdstar^{5/6} \Mdisk^{-1} \Mstar^{-4/3},
\end{align}
where $\ap, \Mp, \rhop$ denote the orbital radius, mass, and density of the planet, respectively, while $\Mstar$ denotes the mass of the central star.

\section{An Approximation of the $\eta$-Dependence in Equation 6}\label{app:fit}

\begin{figure}
\includegraphics[width=\columnwidth]{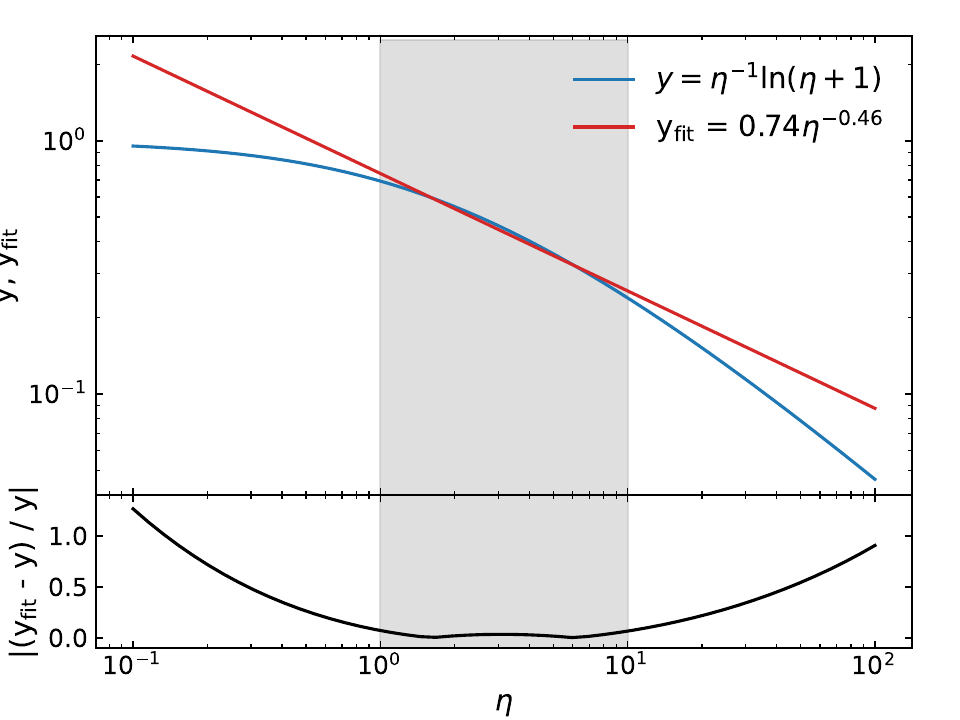}
\caption{
\textit{Top:}$\eta$-Dependence of $\macctot$ (blue curve, from \autoref{eq:m_reacc})) and its power-law approximation, $0.74 \eta^{-0.46}$ (red curve), within the range $1 \lesssim \eta \lesssim 10$ (shaded region).
\textit{Bottom:} Absolute relative error between the exact function and the power-law approximation.
}
\label{appfig:eta_dep_fit}
\end{figure}

From \autoref{eq:m_reacc}, we find that the total mass eventually accreted by the planet is given by $\macctot = \Mdiski \eta^{-1} \ln(1+\eta)$. To express $\macctot$ in terms of familiar parameters such as $\Mp$ and $\fesc$, we approximate the $\eta$ dependence as a power-law in the range $1 \lesssim \eta \lesssim 10$, relevant for Earth-like terrestrial planets with of $\fesc \sim 0.01$ (see \autoref{eq:eta_approx}). We find $\eta^{-1} \ln(1+\eta) \simeq 0.74 \eta^{-0.46}$ (\autoref{appfig:eta_dep_fit}), which gives us the expression of $\macctot$ presented in \autoref{eq:m_reacc_tot_approx}.


\bsp	
\label{lastpage}
\end{document}